\documentclass[onecolumn,amsmath,showkeys,amssymb]{revtex4}
\usepackage{graphicx,color}
\usepackage{amsbsy}
\usepackage{amsmath,amsthm,mathrsfs,amsopn}
\usepackage{verbatim}
\usepackage{color}
\usepackage{afterpage}

\usepackage[framemethod=tikz]{mdframed}  

\DeclareMathOperator{\tr}{tr}
\DeclareMathOperator{\sign}{sign}

\newcommand{\fDgD}{\ensuremath{f_{u_\tau} D_v + g_{v_\tau}D_u}}
\newcommand{\ssup}{\ensuremath{\mathbf{s}_\mathrm{sup}}}
\newcommand{\snonreal}{\ensuremath{\mathbf{s}_\mathrm{nonreal}}}

\newtheorem{theorem}{Theorem}
\newtheorem{remark}{Remark}

\newtheorem{property}{Property}

\graphicspath{{images/}}
\begin{document}

\title{Pattern formation in a two-component reaction-diffusion system with delayed processes on a network}
\author{Julien Petit$^{1,2}$, Malbor Asllani$^1$, Duccio Fanelli$^3$, Ben Lauwens$^2$ and Timoteo Carletti$^1$}
\affiliation{1. naXys, Namur Center for Complex Systems, University of Namur, rempart de la Vierge 8, B 5000 Namur, Belgium\\
2. Department of mathematics, Royal Military Academy, Brussels, Belgium\\3. Universit\`a degli Studi di Firenze, Dipartimento di Fisica e Astronomia, CSDC and INFN Sezione di Firenze, via G. Sansone 1, 50019 Sesto Fiorentino, Italy}

\begin{abstract} 
Reaction-diffusion systems with time-delay defined on complex networks have been studied in the framework of the emergence of Turing instabilities. The use of the Lambert $W$-function allowed us get explicit analytic conditions for the onset of patterns as a function of the main involved parameters, the time-delay, the network topology and the diffusion coefficients. Depending on these parameters, the analysis predicts whether the system will evolve towards a stationary Turing pattern or rather to a wave pattern associated to a Hopf bifurcation. The possible outcomes of the linear analysis overcome the respective limitations of the single-species case with delay, and that of the classical activator-inhibitor variant without delay. Numerical results gained from the Mimura-Murray model support the theoretical approach.
\end{abstract}

\maketitle

\section{Introduction}
\label{sec:intro}

Self-organizing phenomena are widespread in Nature and have been studied for a long time in various domains, be it in physics, chemistry, biology, ecology, neurophysiology,  to name a few~\cite{NicolisPrigogine1977}. Despite the rich literature on the subject, there is still need for understanding, analyzing and predicting their behaviors. 

They are commonly based on local interaction rules which determine the creation and destruction of the entities at every place, upon which a diffusion process determines the migration of the components. For this reason reaction-diffusion systems are a common framework of modeling  such systems~\cite{grindrod1991}. 

In a 1952 article in biomathematics, Turing considered a two-species model of morphogenesis~\cite{turing1952}. For the first time, he established the conditions for a stable spatially homogeneous state, to migrate towards a new heterogeneous, spatially patched, equilibrium under the driving effect of diffusion, at odd with the idea that diffusion is a source of homogeneity. Even though the  explanation for morphogenesis has evolved and now relies more on genetic programming, many actual results are based or inspired form this pioneering work. The so-called Turing instabilities help explain by a simple means the emergence of self-organized collective patterns. 

The geometry of the underlying support where the reaction-diffusion acts, plays a relevant role in the patterned outcome, it can be because of the non flat geometry~\cite{VareaEt1999} (possibly growing)~\cite{Plazaetal2004} or because of its anisotropy~\cite{BusielloEtAl2015}. Pushing to the extreme the discreteness of the space, scholars have considered reaction-diffusion systems on complex networks; reactions occur at each node and then products are displaced across the network using the available links, thus possibly exhibiting  Turing patterns~\cite{NakaoMikhailov2010}. Since then, the latter have been studied on other complex networks supports, for instance multiplex~\cite{asllanietAl2014,Kouvaris2015} and cartesian product networks~\cite{asllanietAl2015}.

On the other hand, time-delays (also called time lags) kept making their way into more and more mathematical models. Time lags come into play in  classical mechanical engineering applications, for load balancing in parallel computing, in traffic flow models and in still more  fields of network theory. Indeed, they cannot be left alone in the understanding of the interactions between neurons in biology, for distributed,  cooperative or  remote control, in using networks of sensors. Roughly speaking, delays are inherent to virtually all  systems where the time needed for  transport, propagation, communication, reaction or decision making cannot be neglected~\cite{Atay2010, MichielsNiculescu2014}. 

The effects of time-delay on stability come under many flavors. In feedback systems for instance, time-delay can induce or help suppress oscillations (see \cite{MichielsNiculescu2014} p. ix and references therein). On the other hand, introducing the delay can be a very reasonable way to improve the models and avoid unnecessary or complex variants of delay-free approaches to refine the match between predictions and observations. In case of reaction-diffusion systems, this challenge was addressed by some previous work related to delay-driven irregular patterns, where Turing or Hopf bifurcations determine the evolution and the ultimate stable state of the system~\cite{SenetAl2009,TianZhang2013, ZhangZang2014} on continuous domains. These studies focused on delay in the reaction kinetics, not in the diffusion part, and is some cases restricting to the small delays assumption.

In this work, the goal is to tackle a time-delay dependent problem with two profound distinctions with the above studies. First, our system evolves on a discrete domain -a network- and second, we do not limit ourselves to the case that only the reactions are delayed. That is, all the processes taking place in the nodes including the triggering of diffusion need some finite amount of time to occur, resulting in a pure-delay setting~\cite{WuEtAl2012,SteurMichielsEtAl2014}. The retarded behavior could be due to inertia, some limiting physical, technological or human factor, correspond to the processing time, or be by design due to a wait-then-act strategy. A first step in this direction has been recently done in~\cite{CarlettiPetitOneSpecies2015} where authors studied a reaction-diffusion system with time-delay on a top of a complex network; it has been shown that even with only one species, Turing-like traveling waves can emerge, but never stationary patterns. Observe that this result improves the classical one by Turing - i.e. reaction-diffusion without delay - for which at least two species are necessary to have patterns, so the outcome of~\cite{CarlettiPetitOneSpecies2015} is due to the presence of the delay term in the diffusion part.

Building on the premises of the one-species case, we consider a two-component pure-delay reaction-diffusion system. Adapting the linear stability analysis to the time-delay setting and elaborating on the condition for  instability by expanding the perturbation on a generalized basis formed by the eigenvectors of the network Laplacian matrix, we are able to characterize the possible onset of Turing instabilities, stationary patterns and traveling waves. To this end we use the scalar Lambert $W$-function which allows to cast analytical results in closed explicit form depending on the main model parameters: time-delay, diffusion coefficients and network topology. Let us observe that time-delayed systems exhibit a larger set of parameters for which Turing instabilities emerge; more precisely a time-delayed system can have Turing waves for sufficiently large time delays, while the same system without delay cannot develop stationary nor oscillatory patterns.

Hence after having determined the conditions for the emergence of patterns in terms of the time-delay, we analyze the role of the network topology, through its spectral properties, and that of  the diffusion coefficients, and determine once again explicit conditions for the emergence of patterns. The analytical results are complemented and confirmed by direct numerical simulations using the prototype Mimura-Murray model.

The paper is organized as follows. In Section~\ref{sec:themodel} we present the framework of the $2$--species reaction--diffusion model with time delay on a complex network. Section~\ref{sec:Mimura-MurrayModel} is devoted to a brief introduction of the Mimura-Murray model that will be subsequently used to check our analytical results. In Section~\ref{sec:StationaryTuringPatterns} we will provide the conditions for the emergence of stationary patterns, involving the time-delay. Next, we devote Section~\ref{sec:Some-critical-value-of-the-time-delay} to the computation of the delay stability margin of the model, that we compare with the small time-delay approximation estimate of Section~\ref{sec:small_time_delay_first_order_u_v}. Sections~\ref{sec:condition_on_the_Laplacian_eigenvalues} and~\ref{sec:perturbation_about_Dv} will  respectively be dedicated to the study of patterns emergence as a function of the network topology and species mobility. In the final  Section~\ref{sec:conc} we will sum up and conclude.

\section{A two-species model on a network with delayed node processes}
\label{sec:themodel}
We consider a two species activator-inhibitor reaction-diffusion system defined on an undirected  network with $n$ nodes and no self-loops. The network is described by its adjacency matrix, $G$ and let $k_i = \sum_{j=1}^{n} G_{ij}$ denote the degree of the $i$--th node. The Laplacian matrix $L$ of the network is defined by $L_{ij} = G_{ij} - k_i \delta_{ij}$. The time-dependent concentrations of the activator, respectively the inhibitor, in node $i$ will be denoted by $u_i(t)$, respectively $v_i(t)$. The reaction of such quantities in each node is modeled via two nonlinear functions $f$ and $g$, describing thus the creation and/or destruction of the activator and inhibitor. We assume the evolution of such processes to be submitted to some constant time-delay $\tau_r>0$. Finally the activator and inhibitor move across the network links; such process is characterized by two diffusion coefficients $D_u>0$ and $D_v>0$ and also by a time-delay $\tau_d>0$ that can be thought to be the result of some processing time in the nodes, the species are submitted to, before crossing the link. The model is thus described by a system of delay differential equations:
\begin{align}
\label{eq:PDS}
\dot{u}_i(t)  &= f(u_i(t-\tau_r),v_i(t-\tau_r) )+ D_u \sum_{j=1}^{n}L_{ij}u_j(t-\tau_d) \nonumber\\
\dot{v}_i(t)  &= g(u_i(t-\tau_r),v_i(t-\tau_r) ) + D_v \sum_{j=1}^{n}L_{ij}v_j(t-\tau_d) \quad \forall i=1,\dots,n\text{ and $t>0$.}
\end{align}
Let us observe that for a sake of simplicity we assumed that the delays are independent from the nodes and the links, moreover  not jeopardizing the possible types of pattern such model can produce, we add the additional constraint that $\tau_r =\tau_d = \tau >0$. The above system is complemented with \lq\lq initial conditions\rq\rq $u_i(t) = \phi_{u,i}(t)$ and $v_i(t) = \phi_{v,i}(t)$, $t\in [-\tau,0]$, the latter being real-valued continuous functions defined on $[-\tau,0]$. In the sequel, we will write $x_\tau(\cdot)$ for $x(\cdot-\tau)$, $x=u,v$, and drop the reference to the node index where there is no risk of ambiguity. 

Our aim is to determine the conditions for the onset of patterns according to a Turing mechanism. The system should exhibit an asymptotically stable equilibrium in absence of the diffusion part and moreover once the latter is tuned on, the equilibrium should become unstable, namely any small perturbation from such equilibrium will demonstrate an exponential growth in the linear regime. This growth will be slowed down thanks to the nonlinearities of the model,  eventually leading to a new steady state, with spatial and possibly also time variations. 

\subsection{Characteristic equation without diffusion}
Let us first consider the system without diffusion and let us denote by
$(u_i,v_i)_{1\leq i\leq n}=(\hat{u},\hat{v})$ the homogeneous equilibrium, namely $f(\hat{u},\hat{v}) = g(\hat{u},\hat{v})=0$. To determine its stability we define the small perturbations $\delta u $=$ u-\hat{u}$ and $\delta v $=$ v - \hat{v}$ and thus by linearizing \eqref{eq:PDS} about $(\hat{u},\hat{v})$ we get (observe that $x_w$ denotes the partial derivate of $x$ with respect to $w$ with $x=f,g$ and $w = u_\tau,v_\tau$, evaluated at $(\hat{u},\hat{v})$):
\begin{equation}
\label{eq:matrixFormSystemWithoutDiffusion}
\begin{pmatrix} \dot{\delta u}\\ \dot{\delta v} \end{pmatrix} = 
\mathbf{J}
\begin{pmatrix} {\delta u_\tau}\\{\delta v_\tau}\end{pmatrix} 
\qquad \mbox{where} \qquad 
\mathbf{J} = 
\begin{pmatrix} f_{u_\tau} & f_{v_\tau}\\g_{u_\tau} & g_{v_\tau}\end{pmatrix}
\end{equation}
with initial conditions $\delta u (t) = \delta \phi_u(t)$,  $\delta v(t) = \delta \phi_v(t)$, $t\in [-\tau,0]$, where again $\delta \phi_u,\delta \phi_v \in \mathcal{C}([-\tau,0],\mathbb{R})$. The existence and uniqueness of the solution to this linear problem of retarded type is guaranteed for any such initial conditions (see \cite{MichielsNiculescu2014}, chapter 1) as one can also  explicitly check using the method of steps. Looking for a solution of the form
$(\delta u,\delta v) = (c_1,c_2)e^{\lambda t}$, from \eqref{eq:matrixFormSystemWithoutDiffusion} we get a nonlinear eigenvalue problem, whose characteristic equation is
\begin{equation}
\label{eq:characNoDIffusion}
\det \Delta(\lambda) = 0 \qquad\mbox{where} \qquad \Delta(\lambda)  =  \lambda I - \mathbf{J} e^{-\lambda \tau} \text{ is the characteristic matrix.}
\end{equation}
The solutions, called characteristic roots, solve the quasi-polynomial equation	
\begin{equation}
\label{eq:quasipolynomialWithoutDiffusion}
\lambda^2 + (-\lambda\tr \mathbf{J} )e^{-\lambda \tau} +\det \mathbf{J} e^{-2\lambda \tau} = 0,
\end{equation}
and generically (nonzero $\mathbf{J}$ and positive delay) there are infinitely many solutions. The null solution of \eqref{eq:matrixFormSystemWithoutDiffusion} is asymptotically stable if the spectral abscissa,
\begin{equation}
\label{eq:spectabs}
\mathbf{s}= \sup_{\lambda \in \mathbb{C}} \left\{ \Re (\lambda) \mid \det  \Delta(\lambda) = 0 \right\},
\end{equation}
is strictly negative. In other words, all characteristic roots of the system without diffusion lie in the open left-half plane. Several numerical methods exist to compute (an approximation of) these roots. Note however that from a stability perspective, only the knowledge of a finite number of roots is relevant. Indeed, a standard result in delay differential equations is that there are finitely many characteristic roots in any right-half plane. 

\begin{figure}[h]
\begin{center}
\includegraphics[width=0.5\textwidth]{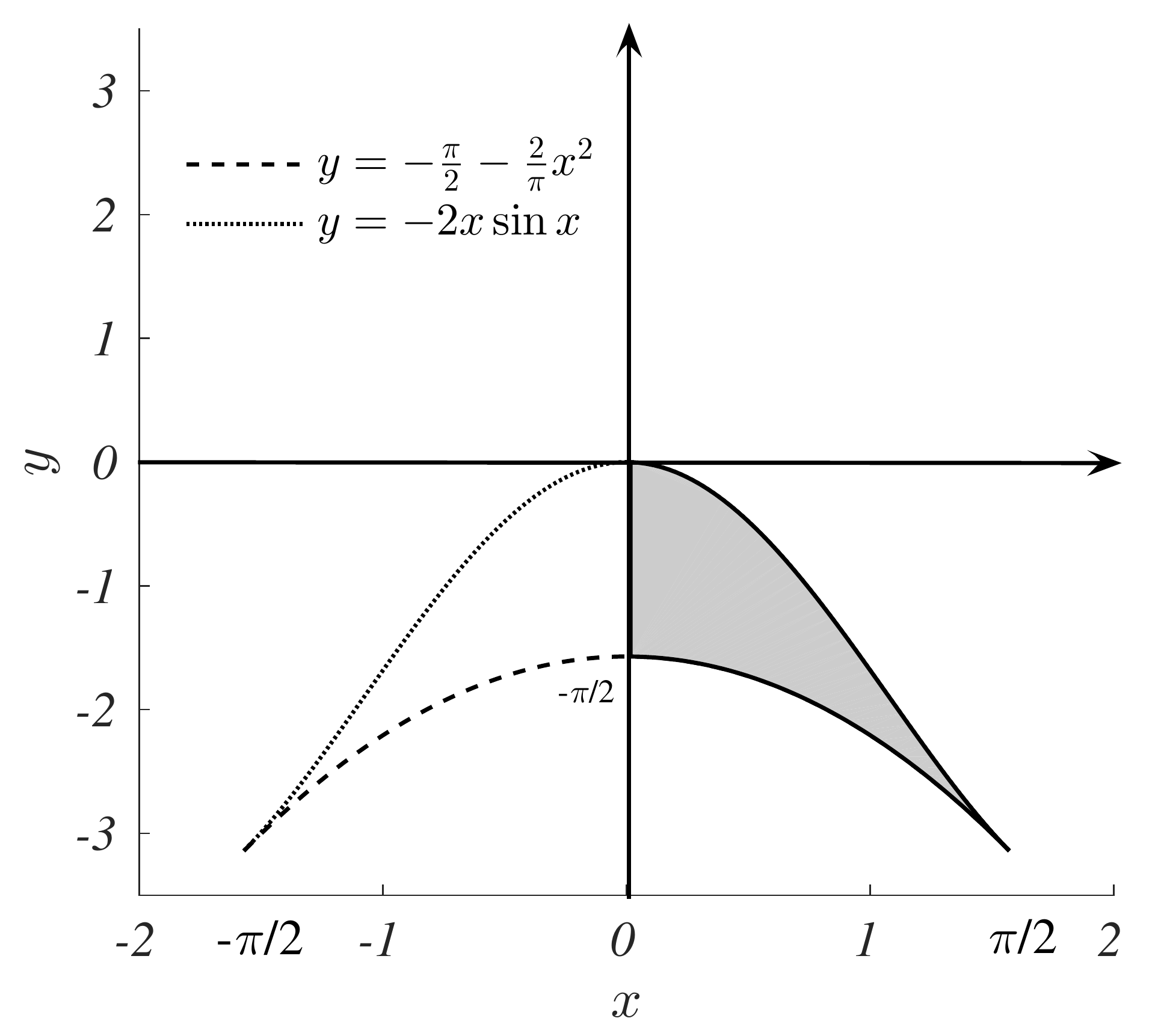} 
\end{center}
\caption{The stability domain given by Eqs.~\eqref{eq:stabdom1} and~\eqref{eq:stabdom2}. The couples $(x,y)=(\tau\sqrt{\det \mathbf{J}},\tau\tr \mathbf{J})$ belonging to the \lq\lq banana--like\rq\rq domain, determine a stable homogeneous equilibrium. Observe also that because of the positivity of $\tau$, such domain is actually restricted to the right half plane (shaded in light grey).}
\label{fig:satbdomnodiff}
\end{figure}

In the present work we will avoid numerical techniques allowing to solve \eqref{eq:quasipolynomialWithoutDiffusion}  and instead rely on an analytical approach. The stability analysis could  be carried out via the following result, first established by \cite{HaraetAl1996}, with \cite{Matsunaga2007} presenting  a simpler proof later on. 
\begin{theorem}
The zero solution of \eqref{eq:matrixFormSystemWithoutDiffusion} is asymptotically stable if and only if 
\begin{equation}
\label{eq:stabdom1}
2 \sqrt{\det \mathbf{J}} \sin (\tau \sqrt{\det \mathbf{J}} )< -\tr \mathbf{J} < \frac{\pi}{2\tau} + \frac{2\tau\det \mathbf{J}}{\pi}
\end{equation}
and 
\begin{equation}
\label{eq:stabdom2}
0 < \tau^2 \det \mathbf{J} < \left(\frac{\pi}{2}\right)^2.
\end{equation}
\end{theorem}
The relevance of this theorem relies on the fact that the statement is expressed in
terms of $\tau$, $\det \mathbf{J}$ and $\tr \mathbf{J}$ and moreover, whenever
one can use the real variables $x=\tau \sqrt{\det \mathbf{J}}$ (that is $\det \mathbf{J}>0$) and $y=\tau\tr
\mathbf{J}$, it allows a simple geometrical check
for stability: the couple $(\tau \sqrt{\det \mathbf{J}},\tau\tr\mathbf{J})$
should lay in the shaded domain reported in the
Fig.~\ref{fig:satbdomnodiff}. However the information obtained remains mainly
qualitative and thus to have a more accurate description of patterns emergence
we have to resort to quantitative methods such as the Lambert $W$--function~\cite{CorlessEtAl1996}.

Several results are available in the literature about the study of the zeros
of Eq.~\eqref{eq:quasipolynomialWithoutDiffusion}, in particular in the case
of commensurable delays. Let us recall the results by \cite{GuetAl2005,NiculescuetAl2005,Jarlebring2009} where authors characterized the position of the zeros relatively to the imaginary axis, the switches and reversals (roots crossing the axis respectively towards instability and stability), just to mention few of them. Some results assume the time lags to be the main parameters, which is not an obvious choice here. Roughly, we can categorize the available approaches into two classes; the time-domain methods are based on Lyapunov-Krasovskii functionals and yield conservative necessary stability conditions. And the spectral approaches having the advantage of yielding exact results, but at the price to be often less tractable. Our choice to use the Lambert $W$-function sets us in the latter class, however we will be able to provide closed-form expressions taking advantage of the particular form of the characteristic equation. 

The transcendental equation \eqref{eq:characNoDIffusion} can be written as  
\begin{equation}
\det \left(  \lambda e^{\lambda \tau}  I - \mathbf{J} \right) = 0\, ,
\end{equation}
left and right multiplying the previous equation by a suitable non singular
matrix $P$ such that $P^{-1}\mathbf{J}P=\left( \begin{smallmatrix}\mu_1 &
    0 \\ * & \mu_2\end{smallmatrix}\right)$, where $*$ denotes any real number
and $\mu_1,\mu_2$ the eigenvalues of $\mathbf{J}$, we can solve Eq.~\eqref{eq:characNoDIffusion} using the Lambert
$W$--function to find the characteristic roots,
$\lambda$, namely to solve $\lambda e^{\lambda \tau}=\mu_i$ for $i=1,2$. In this way the spectral
abscissa~\eqref{eq:spectabs} can be rewritten as 
\begin{equation}
\label{eq:spectabs2}
\mathbf{s}= \sup_{\lambda \in \mathbb{C}} \left\{ \Re (\lambda) \mid \lambda e^{\lambda \tau}=\mu_i\, , i=1,2\right\}=\sup_{\lambda \in \mathbb{C}} \left\{ \Re (\lambda) \mid \lambda =\frac{1}{\tau}W(\mu_i)\, , i=1,2\right\}\, .
\end{equation}

The Lambert $W$--function has an infinite number of branches $W_k$, indexed by $k\in \mathbb{Z}$, among which only $W_{-1}$ and $W_0$ can assume real values. The latter being defined for $z$ real and greater than ${-1}/{e}$, is named the principal branch and it is real valued. Given $z=x+i y$, $w=\xi+i\eta$ and assuming $z=we^w$, that is $w=W_k(z)$ for some $k$, one can relate the variables $(x,y)$ and $(\xi,\eta)$ as follows:\begin{equation}
\label{eq:xyFunctionXiEta}
\begin{cases}
x  =  e^\xi (\xi \cos \eta - \eta \sin \eta)\\
y  =  e^\xi (\eta \cos \eta + \xi \sin \eta )\, .
\end{cases}
\end{equation}

\begin{figure}[h]
\begin{center}
\includegraphics[width=8cm]{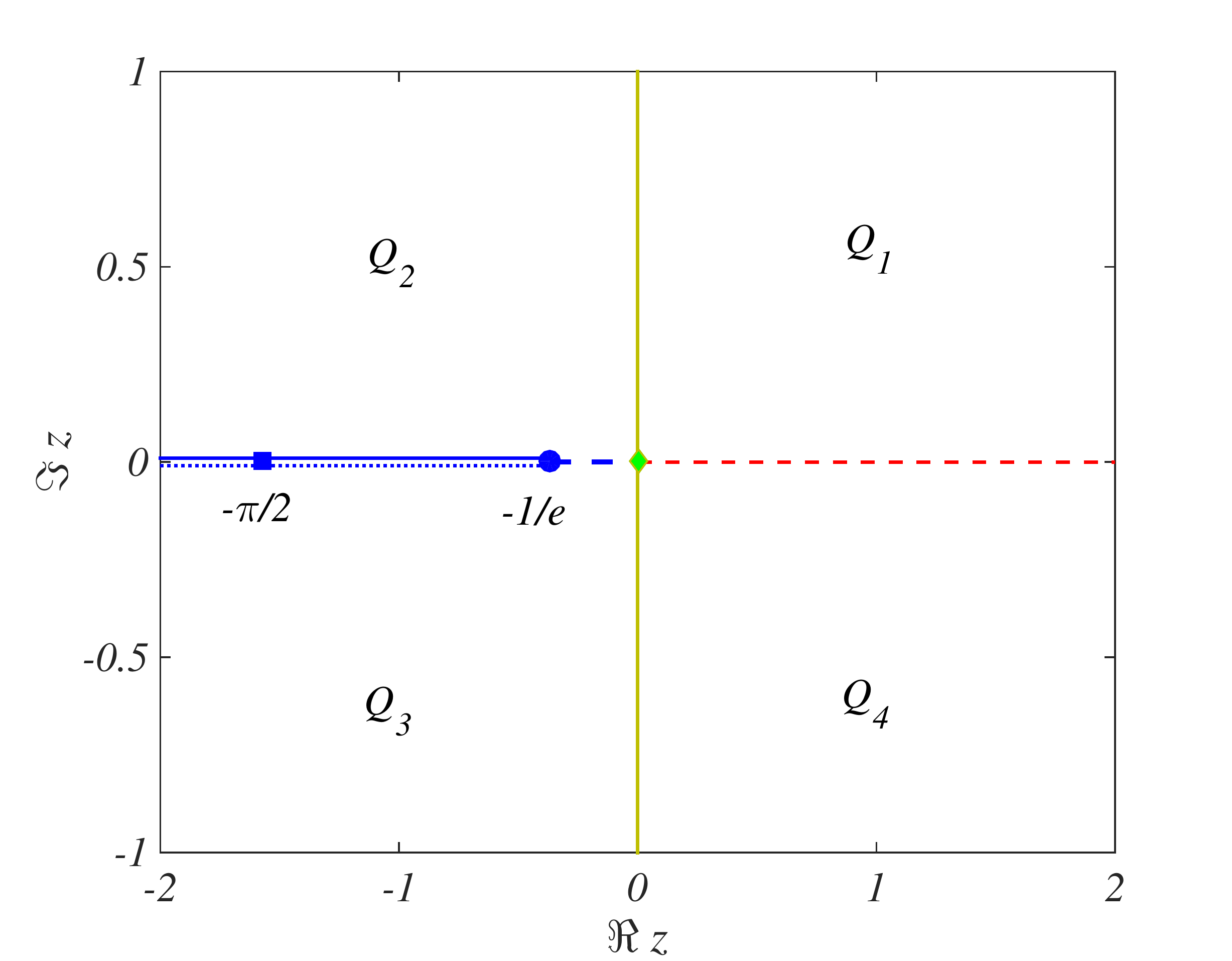}\quad\includegraphics[width=8cm]{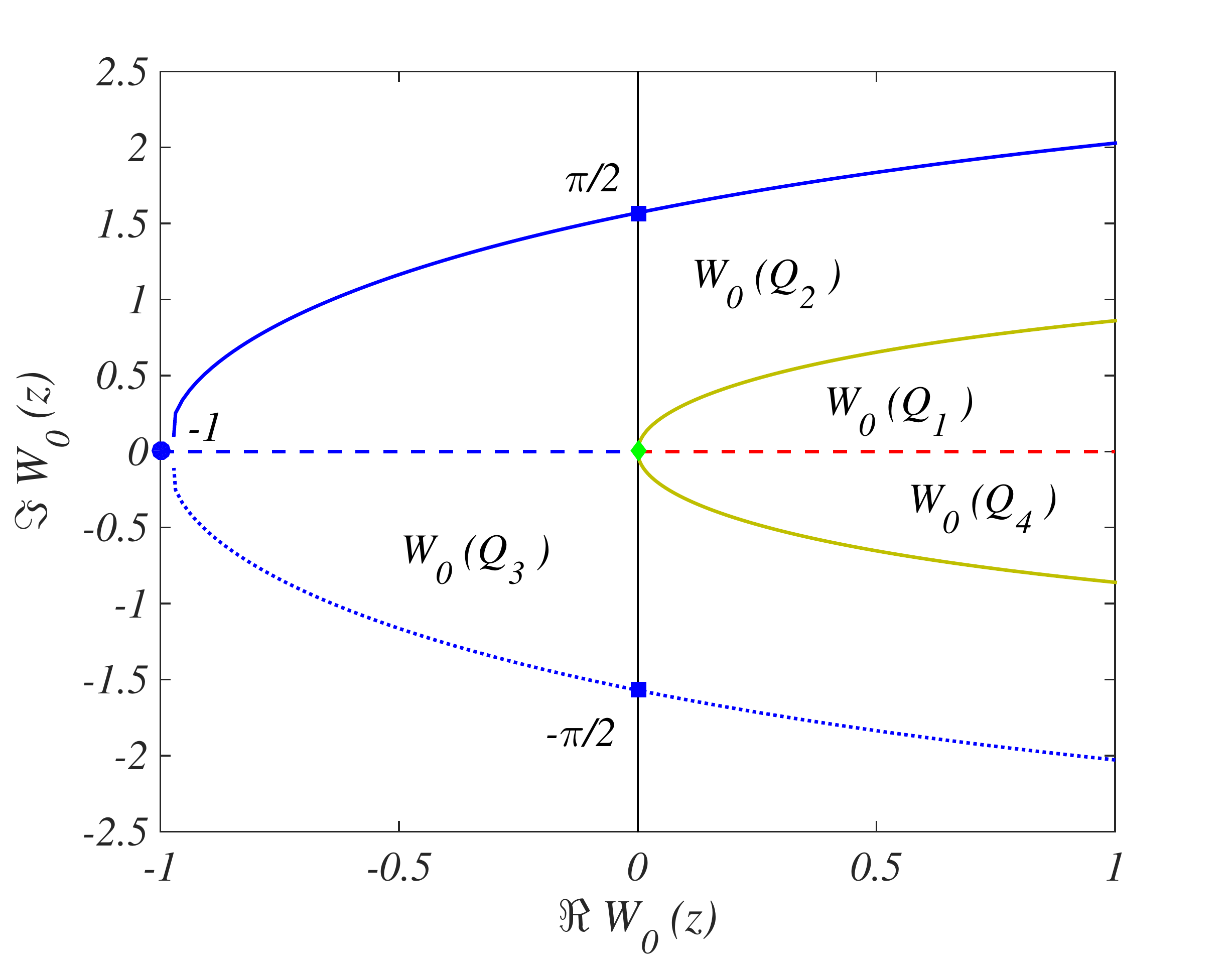}
\end{center}
\caption{The Lambert $W$ function: principal branch $W_0(z)$. Left panel the complex plane $z\in\mathbb{C}$, we represent the upper part of the branch cut $\{z : - \infty < \Re z \leq -e^{-1}\, , \, \Im z=0^+\}$ by a solid line and the lower part of the branch cut $\{z : - \infty < \Re z \leq -e^{-1}\, , \, \Im z=0^-\}$ by a dashed line; the circle denotes the point $(-1/e,0)$ while the square the point $(-\pi/2,0)$. Right panel the complex plane $w\in\mathbb{C}$, where $w=W_0(z)$, the solid blue line is the image of the upper part of the branch cut $\Im w \mapsto \Re w=-\Im w\, \mathrm{co}\!\tan \Im w$ for $0<\Im w <\pi$, while the dashed blue line is the image of the lower branch cut through $W_0$, $\Im w \mapsto \Re w=-\Im w \,\mathrm{co}\!\tan \Im w$ for $-\pi<\Im w <0$. The circle of coordinates $(-1,0)$ is the image of the point $(-1/e,0)$ and the square $(0,\pi/2)$, respectively $(0,-\pi/2)$, is the image of the point $(-\pi/2,0^+)$, respectively $(-\pi/2,0^-)$. The red dashed line is the image of the positive real axis while the green curved line is the image of the imaginary axis $\Re z=0$.}
\label{fig:lambertW}
\end{figure}

Roughly speaking $W_0$ bends the $z$ plane (cut along $\Re z<-e^{-1}$) into a parabolic like domain in the $w$ plane, whose boundary curves $\Im w \mapsto \Re w=-\Im w \,\mathrm{co}\!\tan \Im w$ (blue solid and dotted curves in Fig.~\ref{fig:lambertW}) are bounded by $\pi$ and $-\pi$.

The following property~\cite{ShinozakiMori2006} motivate the use of the Lambert $W$--function for the stability analysis of time delay systems, and will serve throughout this document.\\
\begin{property}
	\label{lem:wkw0}
	For every $z\in \mathbb{C}$, we have
	\begin{equation}
	\label{eq:wkw0}
	\max_{k\in \mathbb{Z}}\Re W_k(z)=\Re W_0(z)\,.
	\end{equation}
\end{property}

It then follows that the spectral abscissa is given by 
\begin{equation}
\label{eq:maxReLambaNoDIffusion}
\mathbf{s} =\frac{1}{\tau} \max\left\{\Re W_0(\tau \mu_1),\Re W_0(\tau \mu_2)\right\}\, .
\end{equation}

\begin{figure}[h]
	\begin{center}
		\includegraphics[width=0.4\textwidth]{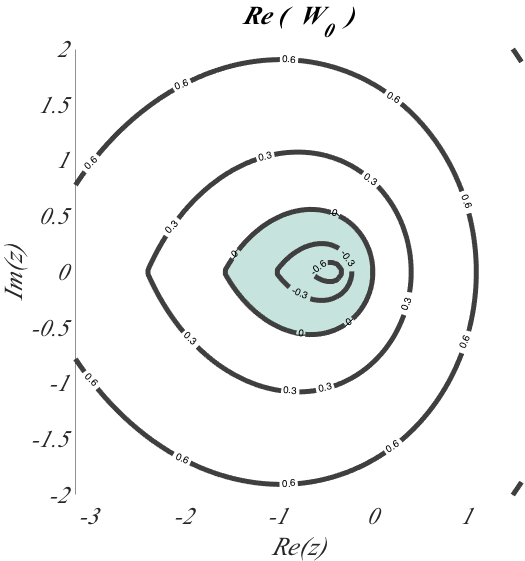} \qquad
		\includegraphics[width=0.4\textwidth]{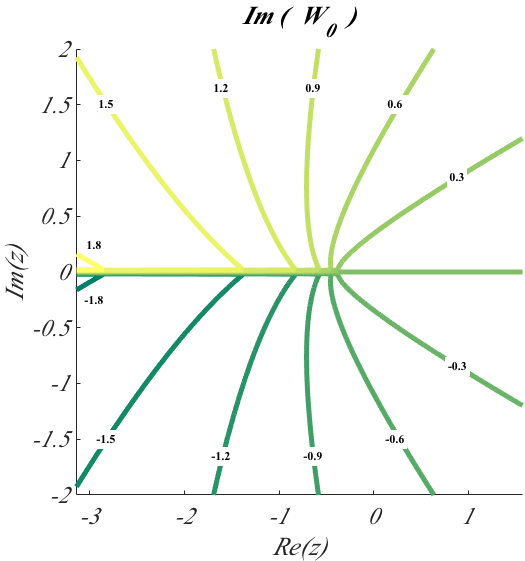} 
	\end{center}
	\caption{Level curves for the principal branch of the Lambert function ($\Re W_0$ left panel and $\Im W_0$ right panel). The shaded area on the left panel corresponds to the interior of the domain with boundary $\Re W_0 = 0$ and represents the stability region for $\tau \mu_1$ and $\tau \mu_2$. Note that the imaginary part is nonzero everywhere except on part of the real axis, for $\Re z \geq -e^{-1}$. It is positive above the real axis, and negative below it.} 
	\label{fig:stabilityRegionNoDiffusion}
\end{figure}

The stability domain of the system without diffusion corresponds thus to all combinations of $\tau\mu_1$ and $ \tau\mu_2$  whose image by $W_0$ is on the left of the imaginary axis, being by definition the time delay to be positive. The boundary of this domain in the complex plane is obtained by letting $\xi = 0$ and let $\eta$ to vary in $[-\frac{\pi}{2},\frac{\pi}{2}]$ in \eqref{eq:xyFunctionXiEta}:
\begin{equation}
\begin{cases}
x & = -\eta \sin \eta \\
y & = \eta \cos \eta
\end{cases}, \qquad -\frac{\pi}{2}\leq\eta\leq\frac{\pi}{2}.
\end{equation}
The stability region is visible in the left panel of Fig.~\ref{fig:stabilityRegionNoDiffusion} where we show the level curves of $\Re W_0(z)$ as a function of $\Re z$ and $\Im z$. 
Based on the above result, the necessary and sufficient condition on the eigenvalues of $\mathbf{J}$ for the stability without diffusion are: 
\begin{equation}
\label{eq:CNSconditionNoDiffusion}
	\frac{-\pi}{2} < \Re (\tau \mu_{1,2}) < 0 \quad \mbox{ and } \quad \left| \Im (\tau \mu_{1,2} )\right| < \hat{\eta} \cos  \hat{\eta}, 
	\end{equation}
with $ \hat{\eta}$ the solution of $\tau \frac{\tr \mathbf{J}}{2} = - \hat{\eta} \sin  \hat{\eta}$ in the interval $(0,\pi/2)$. Note that necessarily, $-\pi <\tau \tr \mathbf{J}=\tau(\mu_1+\mu_2)<0$ and $\det(\mathbf{J}) = \mu_1 \mu_2 >0$.

Because we can decouple the equations for the
characteristic roots, we can avoid the use of the matrix version of the Lambert $W$--function~\cite{AslUlsoy2003}, which, moreover,
lacks~\cite{JarlebringDamm2007,CorlessEtAl2007,GomezMichiels2014} a
property similar to Property~\ref{lem:wkw0}.

\subsection{Characteristic equation with diffusion}
Let us now introduce the diffusion and check if  there are parameters values
for which the homogeneous equilibrium loses its stability, allowing thus the
onset of Turing patterns. To perform a linear stability analysis we will
introduce the eigenbasis of the Laplacian matrix $\{\phi^\alpha : \alpha =
1,\ldots,n\}$, being each eigenvector associated to a topological eigenvalue
$\Lambda^\alpha$; we also assume to order the latter as following
$0=\Lambda^1>\Lambda^2\geq \ldots \geq  \Lambda^n$. Using this basis we can
decompose any small perturbation from the equilibrium, as previously done:
\begin{equation}
\label{eq:generalFormSolution}
\begin{pmatrix}
{\delta u_i(t)}\\
{\delta v_i(t)}
\end{pmatrix} = \sum_{\alpha = 1}^{n} 
\begin{pmatrix}
c_1^\alpha\\
c_2^\alpha
\end{pmatrix}e^{\lambda_\alpha t}\phi_i^\alpha,\qquad(i= 1,\ldots,n)
\end{equation}
where  the constants $c_i^\alpha$ ($i= 1,2$ and $\alpha = 1,\ldots,n$) are determined by the history of the system on $[\tau,0)$, roughly speaking the \lq\lq initial conditions\rq\rq of the delay system. The linearized system~\eqref{eq:PDS} with the delayed diffusion now reads:
\begin{equation}
\begin{pmatrix} \dot{\delta u}\\ \dot{\delta v} \end{pmatrix} = 
\mathbf{J}
\begin{pmatrix} {\delta u_\tau}\\{\delta v_\tau}\end{pmatrix} 
+
\begin{pmatrix} D_u & 0\\0 & D_v\end{pmatrix} 
\begin{pmatrix} \sum_j \mathbf{L}_{ij} \delta u_{j,\tau} \\  \sum_j \mathbf{L}_{ij} \delta v_{j,\tau}   \end{pmatrix}\, ,
\end{equation} 
hence inserting the ansatz~\eqref{eq:generalFormSolution} in the former equation, noticing that $\sum_j L_{ij}\phi_j^\alpha=\Lambda^\alpha \phi_i^\alpha$ and using the orthogonality of the eigenvectors, we get for each mode $\alpha$ ($\alpha = 1,\ldots,n$)
\begin{equation}
\lambda_\alpha \begin{pmatrix} c_1^\alpha \\ c_2^\alpha\end{pmatrix} e^{\lambda_\alpha t} = 
\mathbf{J}  \begin{pmatrix} c_1^\alpha\\c_2^\alpha\end{pmatrix} e^{\lambda_\alpha (t-\tau)}+
\begin{pmatrix}D_u & 0\\ 0 & D_v\end{pmatrix} \Lambda^\alpha
\begin{pmatrix} c_1^\alpha\\c_2^\alpha\end{pmatrix} e^{\lambda_\alpha (t-\tau)},	 
\end{equation}
and finally the characteristic equation
\begin{equation}
\label{eq:characPDS}
\det \Delta^\alpha(\lambda_\alpha) = 0 \qquad\mbox{where}\qquad
\Delta^\alpha (\lambda_\alpha) =  \lambda_\alpha I -\left(\mathbf{J} +\begin{pmatrix}D_u & 0\\ 0 & D_v\end{pmatrix} \Lambda^\alpha \right)e^{-\lambda_\alpha \tau}=\lambda_\alpha I-\mathbf{J^\alpha}e^{-\lambda_\alpha \tau}\, ,
\end{equation}
where $\mathbf{J^\alpha}$ is defined by the latter equality. Let us remark
that one could get a straightforward qualitative information using once again
the variables $x,y$ reported in Fig.~\ref{fig:satbdomnodiff}; for fixed $\tau$ and
reaction terms, we will have in this case a set of $n$ points, one for each
$\alpha$, whose position with respect to the \lq\lq banana--like\rq\rq domain
will determine the stability or instability of the homogeneous equilibrium
with diffusion. Let us stress once again that this information should be
complemented with a more detailed analysis once we are interested in
quantitative estimates, as done in the following.

Let us observe that 
\begin{equation}
\label{eq:trJadetJa}
\begin{array}{rcl}
\tr \mathbf{J^\alpha} & = & \tr \mathbf{J} + (D_u + D_v) \Lambda^\alpha\\
\det \mathbf{J^\alpha} & = & \det \mathbf{J} + (f_{u_\tau} D_v + g_{v_\tau}D_u) \Lambda^\alpha + D_u D_v (\Lambda^\alpha)^2\, ,
\end{array}
\end{equation}
and thus $\tr \mathbf{J^\alpha}<\tr \mathbf{J}<0$ for all $\alpha>1$, being $\Lambda^\alpha<0$.
An analysis similar to the one performed in the previous section, allows to rewrite the spectral abscissa for all $\alpha$ as:
\begin{equation}
\label{eq:spectralabscissaModeAlpha}
\mathbf{s}_\alpha = 
\frac{1}{\tau} \max\left\{\Re W_0(\tau \mu^\alpha_1),\Re W_0(\tau \mu^\alpha_2)\right\}\, ,
\end{equation}
where $\mu_{1,2}^\alpha$  are the eigenvalues of $\mathbf{J^\alpha}$
\begin{equation}
\label{eq:eigmuialpha}
\mu_{1,2}^\alpha=\frac{\tr \mathbf{J^\alpha}\pm\sqrt{(\tr \mathbf{J^\alpha})^2-4\det \mathbf{J^\alpha}}}{2}\, .
\end{equation}
Observe that such eigenvalues can be real or complex conjugate; throughout the paper
we will order the eigenvalues in such a way $\Re 
\mu_1^\alpha \leq \Re \mu_2^\alpha $. We can
thus conclude that Turing patterns do emerge if there exist $\hat{\alpha}>1$
such that $s_{\hat{\alpha}}>0$ being by assumption $\mathbf{s}_1 <0$, i.e. the
homogeneous equilibrium is stable in absence of diffusion.

\section{The Mimura-Murray interaction model}
\label{sec:Mimura-MurrayModel}
For a sake of completeness we will present our results using the Mimura-Murray
model; its nonlinear behavior is quite generic and for this reason it has been widely used as benchmark in the literature. The model reaction terms are given by: 
\begin{align}
f(u,v) &= \left( \frac{a+bu-u^2}{c} -v \right) u\\
g(u,v) &=\left(  u-(1+dv)  \right)v,
\end{align}
where $a,b,c,d$ are positive parameters. Observe however that in the following
the variables $(u,v)$ will be replaced with the delayed ones
$(u_\tau,v_\tau)$. Out of the six equilibrium points (two are always unstable,
and two more are trivial with one of the two species levels being zero), we
select the following:  
\begin{equation}
(\hat{u},\hat{v}) = \left (      1+ \frac{1}{2d}   \left(     bd-c-2d+\sqrt{\varDelta}   \right) ,
\frac{1}{2d^2}   \left(     bd-c-2d+\sqrt{\varDelta}  \right) 
\right)
\end{equation}
with $\varDelta = (bd-c-2d)^2 + 4d^2 (a+b-1)$.

\begin{figure}[h]
	\centering
	\begin{center}
		\includegraphics[width=0.4\textwidth]{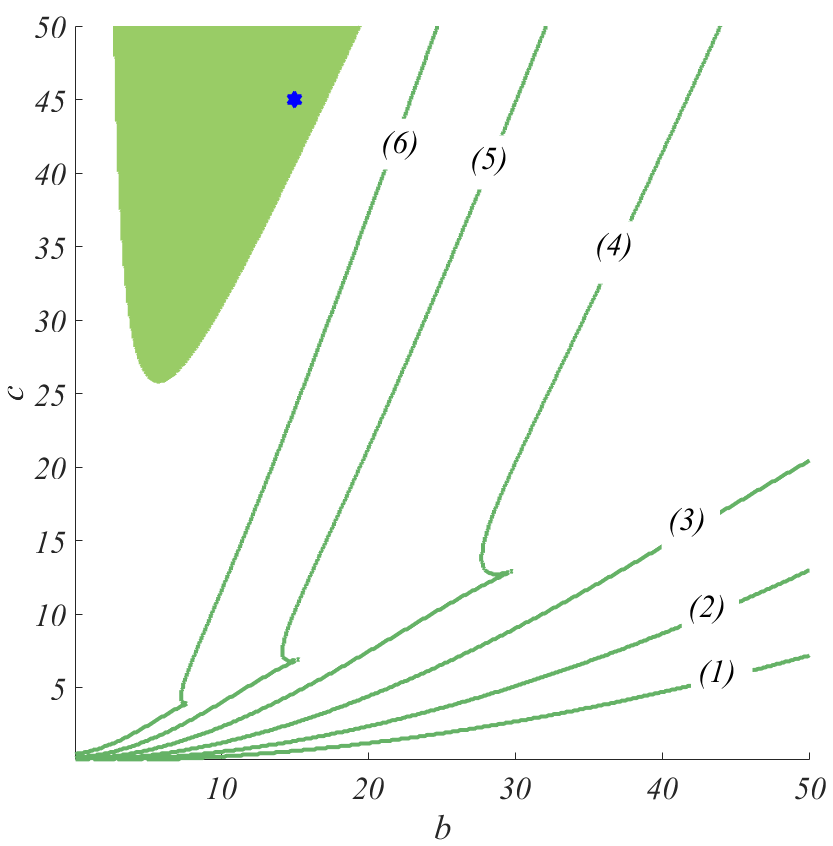} \qquad
		\includegraphics[width=0.4\textwidth]{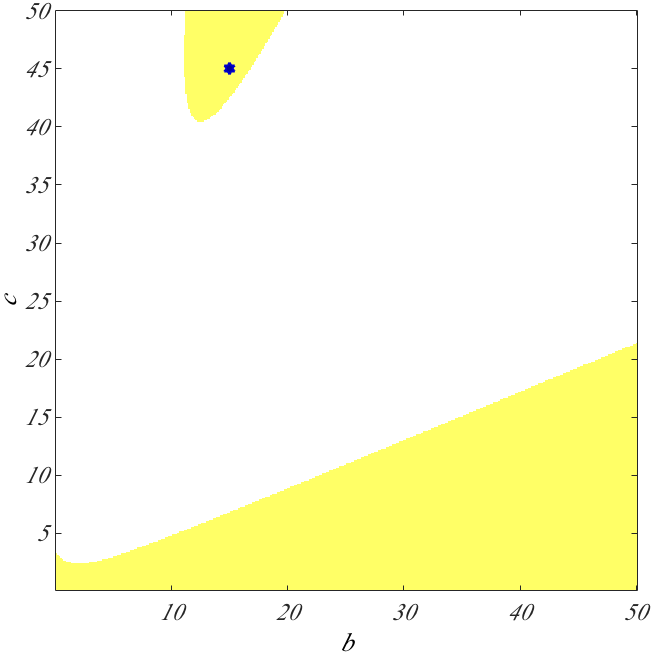} 
	\end{center}
	\caption{Left panel: Stability domain, in the $b-c$ parameters space, of the equilibrium point for the aspatial Mimura-Murray model, for increasing values of the delay: $\tau_1= 0.005$,  $\tau_2 = 0.01$, $\tau_3=0.02$,  $\tau_4=0.04$, $\tau_5=0.08$, $\tau_6= 0.16$. The stability domain related to $\tau_k$ is on the left of each green line labeled by $(k)$, $k=1,\ldots,6$; thus the larger is $\tau$, the smaller is the region for the stability of the homogeneous solution. The green shaded region corresponds to $(b,c)$ values for which the sign of the partial derivatives of $f$ and $g$  are in accordance with the activator role of $u$ and the inhibitor role of $v$. Right panel: character of the eigenvalues. In yellow we report the values of the parameters $(b,c)$ corresponding to real eigenvalues, while in white are the ones associated to complex eigenvalues. The remaining parameters have been set equal to $a = 2.5$ and $d=4.5$. The set of parameters $(a,b,c,d) = (2.5,15,45,4.5)$ used throughout the text is marked by a blue star.} 
	\label{fig:eigenvalJacobian}
\end{figure}

The stability region in the $b-c$ parameter space for the system without
diffusion is reported in the left panel of Fig.~\ref{fig:eigenvalJacobian},
while on the right panel we report the parameters values for which the
eigenvalues of the Jacobian matrix $\mathbf{J}$ are real or complex. 

\begin{remark}
The fact that we consider here a two-species model sets us apart from the one-species version of the pure-delay reaction-diffusion system presented in~ \cite{CarlettiPetitOneSpecies2015} in several aspects. One of these is rather technical, even though it does have its importance in terms of conclusions we draw from the calculations. Namely, the eigenvalues of the Jacobian matrix of the  system are the roots of a quadratic equation, and can be nonreal. Because the characteristic roots and thus the stability properties depend on such eigenvalues, in order not to jeopardize the logical flow upon reading this document, we decided to restrict ourselves to the case of real eigenvalues in the main body and left for the appendix all the details concerning the analytical calculations in the case of nonreal eigenvalues.
\end{remark}

For a sake of concreteness, we select the parameter set $(a,b,c,d) = (2.5,15,45,4.5)$, from which it follows that the homogeneous equilibrium corresponds to $(\hat{u},\hat{v}) = (6.83,1.30)$ and
the eigenvalues $\mu_{1,2}$ are real. This choice is represented by a blue star on Fig.~\ref{fig:eigenvalJacobian}. Of course our results do not depend on this arbitrary choice and the complementary one with complex conjugated eigenvalues could have been used as well.

\section{Stationary Turing patterns}
\label{sec:StationaryTuringPatterns}
In a recent paper~ \cite{CarlettiPetitOneSpecies2015} authors proved that Turing patterns can emerge in a single species reaction-diffusion model on a
network provided the diffusion contains a time delay term; moreover it has
been shown that stationary patterns can never develop in such framework. It is
thus a natural question to investigate if this behavior is proper to
reaction-diffusion model with time delay on networks, endowed with any number
of species. 

To answer this question one should prove that the characteristic root of
Eq.~\eqref{eq:characPDS} with the largest real part is a real and positive
number, avoiding thus the presence of an imaginary part responsible for oscillations. Because of the properties of the Lambert $W$--function and of the definition of spectral abscissa, this amounts to require that the latter is
positive for some $\hat{\alpha}>1$ and moreover the associated characteristic
root $\lambda_{\hat{\alpha}}$ is a real number: $\lambda_{\hat{\alpha}}=\mathbf{s}_{\hat{\alpha}} = 
\frac{1}{\tau} \Re W_0(\tau \mu^{\hat{\alpha}}_2)>0$. On the contrary if $\Im\lambda_{\hat{\alpha}}\neq 0$, and still
$\mathbf{s}_{\hat{\alpha}} >0$, then the presence of a complex part is
responsible for a wave-like behavior. 

The aim of the following part is to
determine the conditions on the eigenvalues $\mu_{i}^\alpha$ to ensure the
previous property to hold, but before to introduce the technical details, let us present the main idea using the following
Fig.~\ref{fig:locationzw}. Here we represent three generic real points,
$a=(1,0)$, $b=(-2,0)$ and $c=(-4,0)$, the first one being
positive and the latter two negative; thanks to the properties of the Lambert
$W$--function, the images through $W_0$ of such points have a nonzero 
  imaginary part and positive real parts provided they are negative enough,
  moreover the real parts are not monotonically ordered, in fact $\Re
  W_0(b)<\Re W_0(a)<\Re W_0(c)$ while $c<b<a$. So
  assuming $a$ corresponds to the real positive largest eigenvalue,
  $\mu_2^{\hat{\alpha}}$ for some $\hat{\alpha}$, then to ensure the
  existence of stationary patterns, one should not have eigenvalues behaving as
  the point $c$, that is $\min_{\alpha}\Re\mu_1^{\alpha}$ should not be too
  much negative. If the eigenvalues $\mu_i^\alpha$ are complex, then things
  can be more involved because their imaginary parts can contribute to the
  real parts of the Lambert $W$--function (see point $b'=(-2,2)$ in
  Fig.~\ref{fig:locationzw}).

\begin{figure}[htbp]
\centering
\includegraphics[width = 0.4\textwidth]{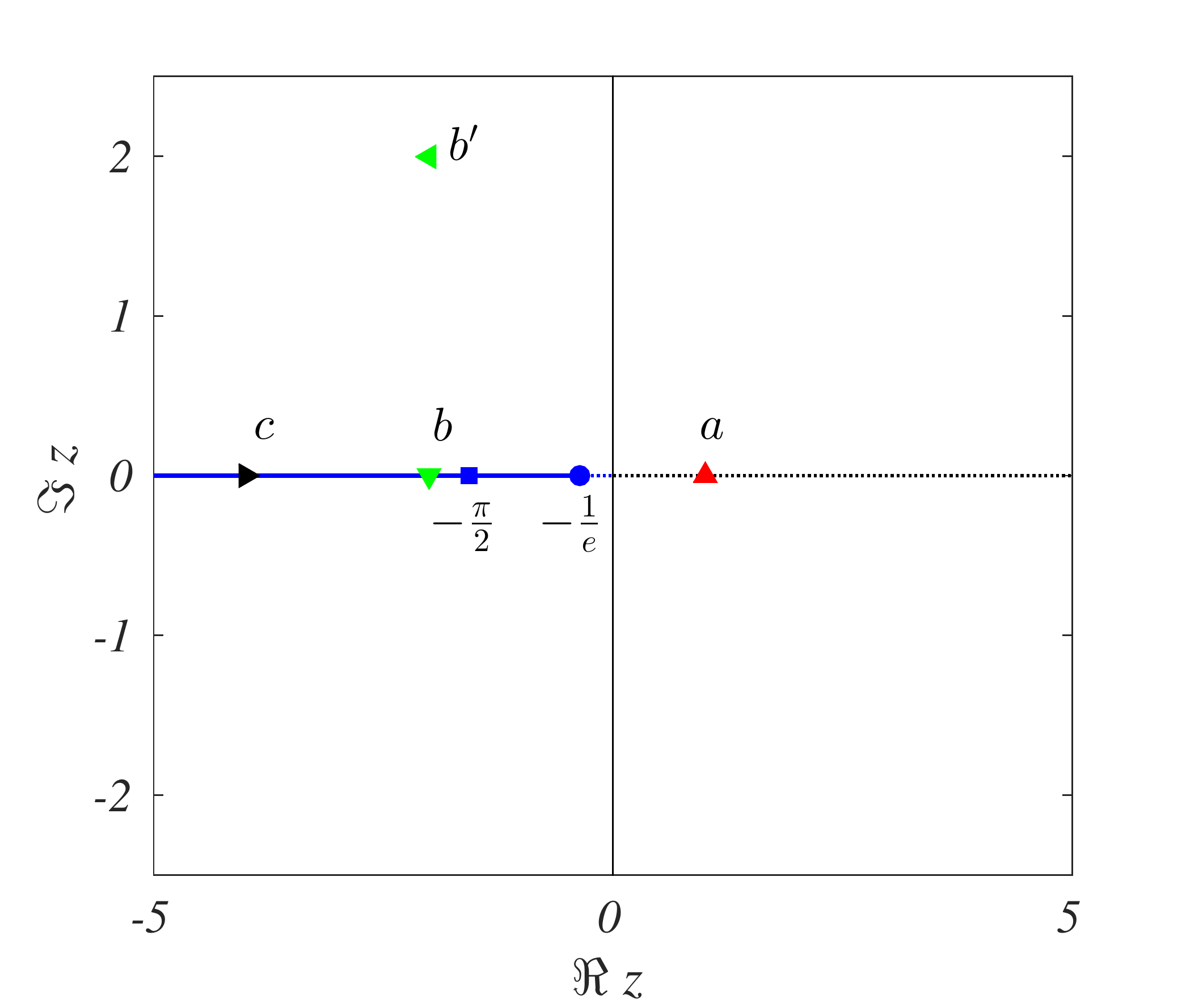} \quad
\includegraphics[width = 0.4\textwidth]{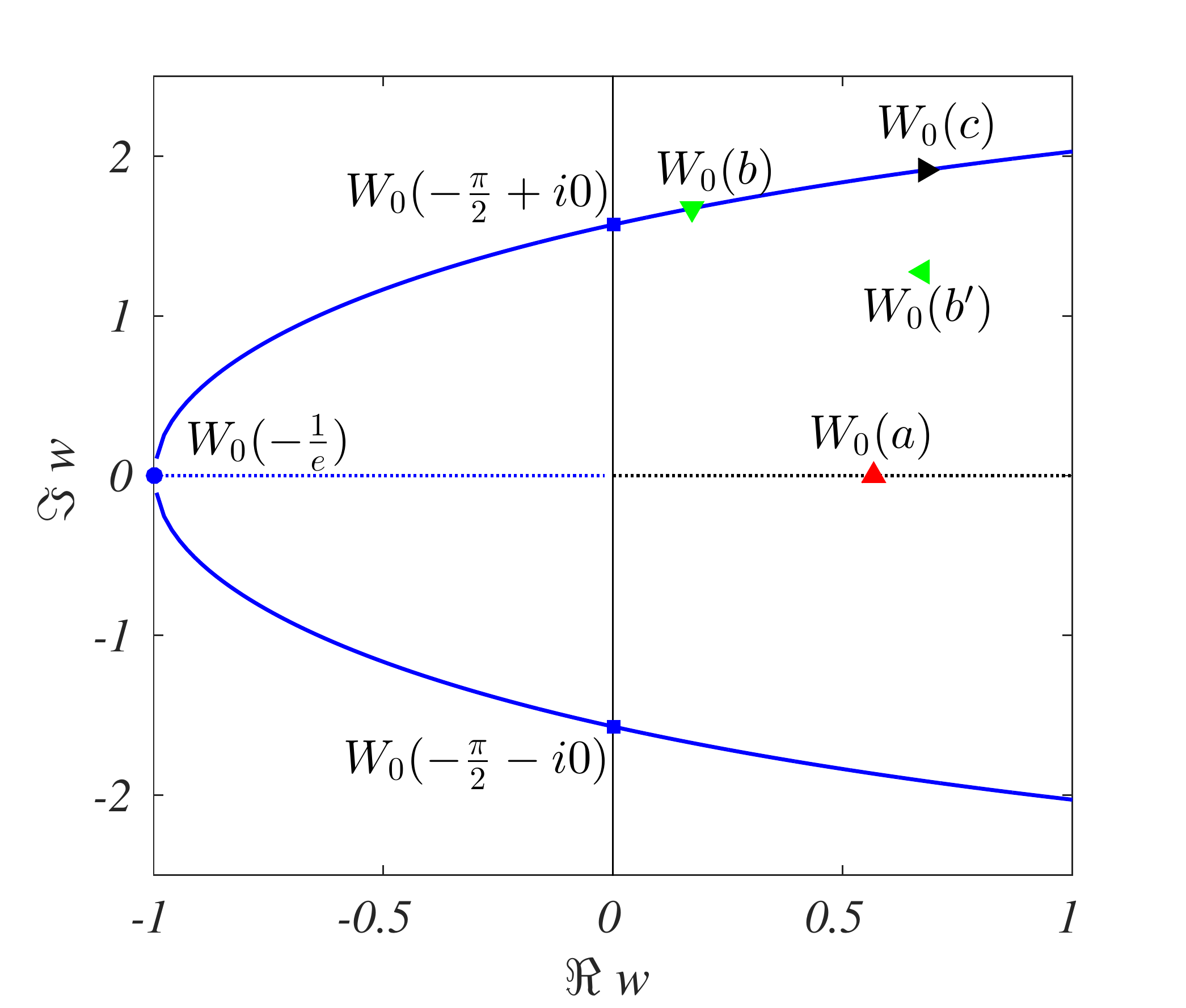} 
\caption{The Lambert $W$--function (principal branch) $z=we^w$,
  $w=W_0(z)$. We show some peculiar points in the $z$--plane (left panel) and
  the corresponding images through $W_0$ (right panel). The circle located at $(−1, 0)$ is the image of the point $(−1/e, 0)$ and the squares
  positioned at $(0,\pi/2)$, respectively at $(0,−\pi/2)$, are the image of the
  point $−\pi/2 +i 0$, respectively $−\pi/2-i0$. The solid blue line (right
  panel) is the image of the upper and lower part of the branch cut $\{z : -
  \infty < \Re z \leq -e^{-1}\, , \, \Im z=0^{\pm}\}$ given by $\Im w
  \mapsto \Re w=-\Im w\, \mathrm{co}\!\tan \Im w$. The black dotted line is
  the image of the positive real axis while the blue dotted line is the image
  of the segment $[-1/e,0]$. $a=(1,0)$ (up red triangle), $b=(-2,0)$ (down
  green triangle) and
  $c=(-4,0)$ (right black red triangle) denote (left panel) three generic real
  points, the first one being positive and the latter 
  negative, we can observe that their images through $W_0$ have a nonzero  imaginary part and positive real parts that are not monotonically ordered $\Re
  W_0(b)<\Re W_0(a)<\Re W_0(c)$ while $c<b<a$. The point $b'=(-2,2)$ (left green  triangle) has the same real part as $b$ but a nonzero imaginary part, we
  can observe that its image (right panel) through $W_0$ has a real part
  larger than $\Re W_0(b)$ but also $\Re W_0(a)$.}
\label{fig:locationzw}
\end{figure}

Starting from the assumption of the stability of the homogeneous equilibrium, in the rest of the section we will thus successively determine the conditions to have a real and positive eigenvalue (section~\ref{sec:existenceOfStationaryModes}), then get an estimate for the associated characteristic root (section~\ref{sec:RHS}) and finally obtain an estimate for the characteristic root associated to the eigenvalues with the smallest real part (section~\ref{sec:LHS}). Section~\ref{sec:SOP} will contain the conclusion of this part.

\subsection{Existence of positive real characteristic roots}
\label{sec:existenceOfStationaryModes}

By the very first definition of the eigenvalues of $\mathbf{J^\alpha}$ given by Eq.~\eqref{eq:eigmuialpha}, they are real and the largest one is positive if and only if:
\begin{equation}
\begin{cases}
(\tr \mathbf{J^\alpha})^2-4\det \mathbf{J^\alpha} >0\\
\tr \mathbf{J^\alpha} +\sqrt{(\tr \mathbf{J^\alpha})^2-4\det \mathbf{J^\alpha}} >0
\end{cases}
\qquad \mbox{for some} \qquad 1<\alpha \leq n\, . 
\end{equation}
By the stability assumption of the system without diffusion, we get $\tr \mathbf{J}<0$ and using the fact that $\tr \mathbf{J^\alpha}<\tr \mathbf{J}$, the previous condition amounts to 
\begin{equation}
\label{eq:condForStatModes}
\begin{cases}
(\tr \mathbf{J^\alpha})^2-4\det \mathbf{J^\alpha} >0\\
\det \mathbf{J^\alpha} <0
\end{cases}
\end{equation}
for some $\alpha=2,\dots,n$. Observe that this is the same set of  conditions as in the case of two species diffusing on a network without delay \cite{NakaoMikhailov2010}, except that here the derivatives are performed with respect to the delayed variables $u_\tau$ and $v_\tau$. The first inequality of Eq.~\eqref{eq:condForStatModes} holds true whenever the second inequality does, so elaborating on the latter we get
\begin{equation}
\det \mathbf{J} + (f_{u_\tau} D_v + g_{v_\tau}D_u) \Lambda^\alpha + D_u D_v (\Lambda^\alpha)^2<0,
\end{equation}
that determines a range for $\Lambda^\alpha$:
\begin{equation}
\label{eq:rangeLambda}
 \Lambda^\alpha \in\left(  \frac{-(\fDgD)-\sqrt{\Delta}}{2D_u D_v},\frac{-(\fDgD)+\sqrt{\Delta}}{2D_u D_v} \right)\, ,
\end{equation}
where $\Delta = (\fDgD)^2-4D_uD_v \det \mathbf{J} $. Since the Laplacian eigenvalues are negative, $\Lambda^\alpha<0$, the condition~\eqref{eq:rangeLambda} can be verified only if $\fDgD>0$ and $\Delta >0$. Because of the assumptions on the roles of activator and inhibitor of $u$ and $v$,  we assume $f_{u_\tau} >0$ and $g_{v_\tau}<0$. Recalling $f_{u_\tau}+g_{v_\tau}=\tr \mathbf{J}<0$, we have 
$\left|\frac{f_{u_\tau}}{g_{v_\tau}}\right|<1$, and so
\begin{equation}
\fDgD >0 \quad \iff  \quad 1>\left|\frac{f_{u_\tau}}{g_{v_\tau}}\right|>\frac{D_u}{D_v}.
\end{equation}
We thus obtained the same condition, $D_v>D_u$, one can find in the framework without delay, but now the working hypothesis is on the sign of derivatives $f_{u_\tau} $ and $g_{v_\tau}$ related to the delayed concentrations of the activator and inhibitor. 

We can determine a critical value for the ratio of the diffusion coefficients $D  =\frac{D_v}{D_u}$, for which the interval given by~\eqref{eq:rangeLambda} reduces to a single point, the condition being $\Delta=0$, namely
\begin{equation}
\label{eq:DeltaIsZero}
f_{u_\tau}^2 D^2 + (2 f_{u_\tau} g_{v_\tau}-4\det \mathbf{J})D+g_{v_\tau}^2 = 0\, .
\end{equation}
The largest root of the latter equation determines such critical value: 
\begin{equation}
\label{eq:Dcrit}
D_\mathrm{crit} = \frac{2\det \mathbf{J}-f_{u_\tau}g_{v_\tau}+2\sqrt{\det \mathbf{J} (\det \mathbf{J}-f_{u_\tau}g_{v_\tau})}}{(f_{u_\tau})^2}\, ,
\end{equation}
let us observe that this root is real and positive ($\det\mathbf{J}>0$ because the homogeneous equilibrium is stable and $f_{u_\tau}g_{v_\tau}<0$ because of the roles of activator and inhibitor of $u$ and $v$). 

Combining the previous equation with \eqref{eq:rangeLambda} gives the corresponding critical value for $\Lambda^\alpha$ for a given $D$. If $D = D_{\mathrm{crit}}$, the rightmost eigenvalue  $\mu_2^\alpha$ has zero value~\cite{NakaoMikhailov2010}. If $D>D_{\mathrm{crit}}$, the system with diffusion is unstable. 

In the following we will be interested in determining the maximum value of the real and positive eigenvalue $\mu_2^\alpha$, as a function of $\Lambda^\alpha$. To achieve this let us compute the derivative of $\mu_2^\alpha$ with respect to $\Lambda^\alpha$ and equal it to zero:
\begin{equation}
\label{eq:derivativeRightSide}
\frac{\mathrm{d}}{\mathrm{d}\Lambda^\alpha} \left(\tr \mathbf{J^\alpha} + \sqrt{(\tr \mathbf{J^\alpha})^2-4\det \mathbf{J^\alpha}}\right)= 0 \iff
 \sqrt{(\tr \mathbf{J^\alpha})^2-4\det \mathbf{J^\alpha}} = -\tr \mathbf{J^\alpha} + 2 \frac{\mathrm{d}(\det \mathbf{J^\alpha})/\mathrm{d}\Lambda^\alpha}{\mathrm{d}(\tr \mathbf{J^\alpha})/\mathrm{d}\Lambda^\alpha}\, . \end{equation}
Using the definitions~\eqref{eq:trJadetJa} we get
\begin{equation}
\label{eq:Fprim}
\frac{\mathrm{d}\tr \mathbf{J^\alpha}}{\mathrm{d}\Lambda^\alpha} = D_u + D_v \quad \text{ and } \quad
\frac{\mathrm{d}\det \mathbf{J^\alpha}}{\mathrm{d}\Lambda^\alpha} = \fDgD + 2D_uD_v\Lambda^\alpha\, ,
\end{equation}
and inserting the above into Eq.~\eqref{eq:derivativeRightSide} we finally obtain
\begin{equation}
\label{eq:forLambda}
c_2(\Lambda^\alpha)^2 +c_1\Lambda^\alpha+c_0 = 0\, ,
\end{equation}
where
\begin{eqnarray}
c_2&=&-D_uD_v(D_u-D_v)^2\\
c_1&=&(4D_uD_v(\fDgD)-2D_uD_v\tr \mathbf{J}(D_u+D_v))\notag\\
c_0&=&\det \mathbf{J} (D_u+D_v)^2 + (\fDgD)^2 - \tr \mathbf{J} (\fDgD)(D_u+D_v)\notag\, .
\end{eqnarray}
Observe that $c_2$ is negative, $c_1$ is positive because by hypothesis $\fDgD>0$ and $\tr \mathbf{J}<0$, and $c_0>0$ because $\det \mathbf{J}>0$, $\fDgD>0$ and $\tr \mathbf{J}<0$; this implies that Eq.~\eqref{eq:forLambda} has two real roots, one positive and one negative, the latter hereby named $\Lambda_\mathrm{sup}$. It is given by 
\begin{equation}
\label{eq:Lambda_sta}
\Lambda_\mathrm{sup} =  \frac{-c_1 + \sqrt{c_1^2-4c_2c_1}}{2c_2}\, ,
\end{equation} 
and determines the largest real and positive eigenvalue $\mu_2^{sup}=\mu_2^{\hat{\alpha}}$, where the index ${\hat{\alpha}}$ is such that $\Lambda^{\hat{\alpha}}=\Lambda_\mathrm{sup}$, assuming the Laplacian spectrum to be well approximated by a continuum, otherwise we should write $\mu_2^{sup} = \max (\mu_2^{\hat{\alpha}}, \mu_2^{\hat{\beta}})$, where $\hat{\alpha} < \hat{\beta}$ are the two closest indexes such that $\Lambda^{\hat{\alpha}}\geq \Lambda_\mathrm{sup} \geq \Lambda^{\hat{\beta}}$.

\subsection{Rightmost real positive characteristic root}
\label{sec:RHS}
Now  we have proved the existence of a real and positive eigenvalue $\mu^\alpha_2=\frac{\tr \mathbf{J^\alpha} + \sqrt{(\tr \mathbf{J^\alpha})^2-4\det \mathbf{J^\alpha}}}{2}>0$, we are interested in determining the associated real and positive characteristic root, $\lambda_\alpha$:
\begin{equation*}
\lambda_\alpha=\frac{1}{\tau} \Re W_0\left(\tau\frac{\tr \mathbf{J^\alpha} + \sqrt{(\tr \mathbf{J^\alpha})^2-4\det \mathbf{J^\alpha}}}{2}\right)\, ,
\end{equation*}
and in particular the largest one :
\begin{equation}
\label{eq:rightmostRealCharacRoot}
\ssup=\max_{\Lambda^\alpha}\lambda_{\alpha}=\frac{1}{\tau}  \max_{\Lambda^\alpha} \Re W_0\left(\tau\frac{\tr \mathbf{J^\alpha} + \sqrt{(\tr \mathbf{J^\alpha})^2-4\det \mathbf{J^\alpha}}}{2}\right)\, .
\end{equation}

Observe that over the interval $[-\frac{1}{e},\infty)$, the function $W_0$ is an increasing real valued function, hence we can rewrite: 
\begin{equation}
\ssup= \frac{1}{\tau}W_0\left(\frac{\tau}{2}\max_{\Lambda^\alpha} \left(\tr \mathbf{J^\alpha} + \sqrt{(\tr \mathbf{J^\alpha})^2-4\det \mathbf{J^\alpha}}\right)\right)\, ,
\end{equation}
and using the computation done in the latter section about the derivative of $\mu^{\alpha}_{1,2}$ with respect to $\Lambda^{\alpha}$, we get
\begin{equation}
\ssup= \frac{1}{\tau}W_0\left({\tau}\mu_2^{sup}\right)=\frac{1}{\tau}W_0\left[\frac{\tau}{2}\left(\tr \mathbf{J}^{\hat{\alpha}} + \sqrt{(\tr \mathbf{J}^{\hat{\alpha}})^2-4\det \mathbf{J}^{\hat{\alpha}}}\right)\right]\, ,
\end{equation}
where the index ${\hat{\alpha}}$ is such that $\Lambda^{\hat{\alpha}}=\Lambda_{sup}$ (see Eq.~\eqref{eq:Lambda_sta} and the associated discussion).

\subsection{Rightmost nonreal characteristic root}
\label{sec:LHS}

Let us now consider modes corresponding to nonreal characteristic roots, and thus possibly yielding oscillations. Recalling the properties of the $W$--function, the latter should be associated to eigenvalues $\mu_1^{\alpha}$ with very negative real parts. The dominant one is thus given by:
\begin{equation}
\label{eq:rightmostNonrealCharacRoot}
\snonreal=\frac{1}{\tau}  \max_{\Lambda^\alpha} \Re W_0\left(\tau \mu_1^{\alpha}\right)=\frac{1}{\tau}  \Re W_0\left( \tau\min_{\Lambda^\alpha}\frac{\tr \mathbf{J^\alpha} - \sqrt{(\tr \mathbf{J^\alpha})^2-4\det \mathbf{J^\alpha}}}{2}\right)\, .
\end{equation}
Observe that the above statement holds true only once we restrict ourselves to real eigenvalues.

For a sake a simplicity we hereby limit ourselves to present the results and we invite the interested reader to consult the section~\ref{sec:computationOfStrongestOscillatoryMode} to have more details. One can prove that for $D_u < D_v$ the maximum in Eq.~\eqref{eq:rightmostNonrealCharacRoot} is achieved for $\alpha = n$ if $\mathbf{J}$ has real eigenvalues $\mu_{1,2}\in\mathbb{R}$, as we are now assuming, namely
\begin{equation}
\label{eq:lambdacmplx1}
\snonreal=\frac{1}{\tau} \Re W_0\left(\tau\frac{\tr \mathbf{J}^n - \sqrt{(\tr \mathbf{J}^n)^2-4\det \mathbf{J}^n}}{2}\right)\, .
\end{equation}

\subsection{Stationary or oscillatory patterns}
\label{sec:SOP}

We are now able to conclude this part and state our main result. Stationary patterns should be observed if
\begin{equation}
\label{eq:statvsoscpatt}
\snonreal<\ssup\, ,
\end{equation}
in fact the characteristic roots satisfy the layout given in Fig.~\ref{fig:locationzw} with $W_0(a)=\ssup$ and $W_0(b)=\snonreal$.

Stated differently, given a real and positive $\ssup$ we can find a real and negative number $r_{osc}$ such that $\Re W_0(r_{osc})/\tau=\ssup$;
if for $\alpha=1,\dots,n$ we have $\Re \tau \mu_1^\alpha>r_{osc}$, then the system may  exhibit stationary patterns. This condition is also sufficient if $\mu_1^\alpha$ is real for all $\alpha$; whereas if $\mu_1^\alpha$ is complex for some $\alpha$, the above condition on its real part, $\Re \tau \mu_1^\alpha>r_{osc}$, is not enough to guarantee that $\Re W_0(\tau \mu_1^\alpha)/\tau$ is smaller than $\ssup$, in fact the imaginary part of $\mu_1^\alpha$ can contribute to $\Re W_0(\tau \mu_1^\alpha)/\tau$ and thus making it larger than $\ssup$ (see point $b'$ in Fig.~\ref{fig:locationzw}).

\section{A delay stability margin}
\label{sec:Some-critical-value-of-the-time-delay}

The aim of this section is to analyse the impact of the time-delay on the emergence of the Turing instability. More precisely, suppose the system we are dealing with doesn't involve any delay and it does not have Turing instabilities, we want determine for which value of the delay the system is capable to exhibit Turing patterns. We hence have to assume that the system possesses a stable homogeneous equilibrium when there is no delay, the equilibrium still remains stable with delay and no diffusion, and eventually the diffusion is able to destabilize it (still with the delay).

We know (see for instance Theorem 1.16 in~\cite{MichielsNiculescu2014}) that if the delay is continuously varied in $\mathbb{R}^+$, then loss or acquisition of stability of the zero solution of system~\eqref{eq:matrixFormSystemWithoutDiffusion}, or its version with diffusion, can only come from characteristic roots crossing the imaginary axis excluding the origin. In other words a system where Turing patterns cannot develop without delay, can exhibit waves-like instability for large enough delay, while a system displaying stationary Turing patterns once $\tau=0$ can evolve to a stationary or oscillatory instability once $\tau>0$. Acting as a scaling factor, the time-delay can force the eigenvalues $\mu_{1,2}$ or $\mu_{1,2}^\alpha$  to enter into or to exit form the stability domain of Fig.~\ref{fig:stabilityRegionNoDiffusion} and thus triggers or impedes oscillatory modes to develop. 

The problem of finding the critical time-delays corresponding to crossings, and their associated  characteristic roots $\lambda = j \omega$ for a linear system of DDE's like~\eqref{eq:matrixFormSystemWithoutDiffusion} has been addressed in the literature in a even more general settings, e.g. taking into account multiple delays or in higher dimensions. Following~\cite{Jarlebring2009}, this problem amounts to solve the quasi-polynomial eigenvalue problem
\begin{equation*}
	\lambda^2 + (-\lambda\tr \mathbf{J} )e^{-\lambda \tau} +\det \mathbf{J} e^{-2\lambda \tau} = 0\, .
\end{equation*}
Using a geometrical approach~\cite{GuetAl2005} one can rewrite the previous equation as
\begin{equation}
	1+ a_1(\lambda)e^{-\lambda \tau} +a_2(\lambda)e^{-2\lambda \tau} = 0\, ,
\end{equation}
where $a_1(\lambda) = -\tr \mathbf{J} /\lambda$ and $a_2(\lambda) = \det \mathbf{J}/ \lambda^{2} $. The bifurcation points of the characteristic roots correspond to $\lambda$ crossing the imaginary axis, $\lambda = \pm j \omega$ for some $\omega \in \mathbb{R}_0^+$: 
\begin{equation}
	1+ a_1(j\omega)e^{-j\omega\tau} +a_2(j\omega)e^{-2j \omega \tau} = 0\, ,
\end{equation}
and back to the original form
\begin{equation}
	1-\frac{\tr \mathbf{J}}{\omega} e^{-j(\omega \tau + \pi /2)} - \frac{\det \mathbf{J}}{\omega^2} e^{-2j\omega\tau}=0\, .
\end{equation}
Each one of the three terms in the left-hand can be though as a vector in the complex plane, which are to form a triangle because their sum is zero. Solving the triangle then allows to find the delays and the corresponding crossing frequencies $\omega$. 

Let us observe that we rather prefer to rely our analysis on the use of the Lambert $W$-function and find an expression for the critical delay based on a crossing of the boundary of the stability domain. We will see that once the delay is increased and reaches this critical delay value, further increasing the delay will never bring back stability. In order words, the smallest value of the time-delay corresponding to a crossing of the imaginary axis is the (stability) delay margin of the system. Following the same principe as before, the computation of the delay margin in the case of nonreal-eigenvalues of the Jacobian matrix $\mathbf{J}$ is left to section~\ref{sec:DelayMarginComplexCase}, while we will hereafter consider only the case of real eigenvalues.

The first step is to determine a range of {$\tau-$values} for which the system has a stable homogenous fixed point. As stated above, we assume stability without delay ($\tr \mathbf{J} < 0, \det \mathbf{J} >0$). If the eigenvalues of $\mathbf{J}$ are real numbers, $\mu_{1,2} \in \mathbb{R}$, then stability is obtained by imposing $-\frac{\pi}{2} < \tau  \mu_i < 0$ for $i=1,2$, hence
\begin{equation}
	\tau <  -\frac{\pi}{2 \mu_1} = \frac{-\pi}{\tr \mathbf{J} -\big( (\tr \mathbf{J})^2-4 \det \mathbf{J}\big)^{1/2}}\, . 
\end{equation}

The second step is to add the diffusion which formally  accounts to replace $\tr \mathbf{J}$ and $\det \mathbf{J}$ by $\tr \mathbf{J^\alpha}$ and $\det \mathbf{J^\alpha}$ in the previous analysis. Thus mode $\alpha$ cannot induce an oscillatory behavior if

\begin{equation}
	\label{eq:exact-critical-value-of-time-delay}
	\tau < -\frac{\pi}{2 \mu_1^\alpha}  = \frac{-\pi}{\tr \mathbf{J^\alpha} - \big(  (\tr \mathbf{J^\alpha})^2 - 4 \det \mathbf{J^\alpha} \big)^{1/2}}
\end{equation}

Observe that for modes related to real eigenvalues we have $\frac{\mathrm{d} \mu_1^\alpha }{\mathrm{d}\Lambda^\alpha} >0$ (see Eq.~\eqref{eq:signDerivativeLambda1alpha}), so the condition \eqref{eq:exact-critical-value-of-time-delay} needs be checked only for the greatest  $\alpha$ such that $\mu_1^\alpha \in \mathbb{R}$. So in the case all $\mu_{1}^\alpha$ are real, the critical value of the delay for the system, i.e. the stability margin,  is readily given by $\tau_{\mathrm{crit}} =- \frac{\pi}{2 \mu_1^n}$.

\begin{figure}[htbp]
	\centering
	\includegraphics[width= 1.0\textwidth]{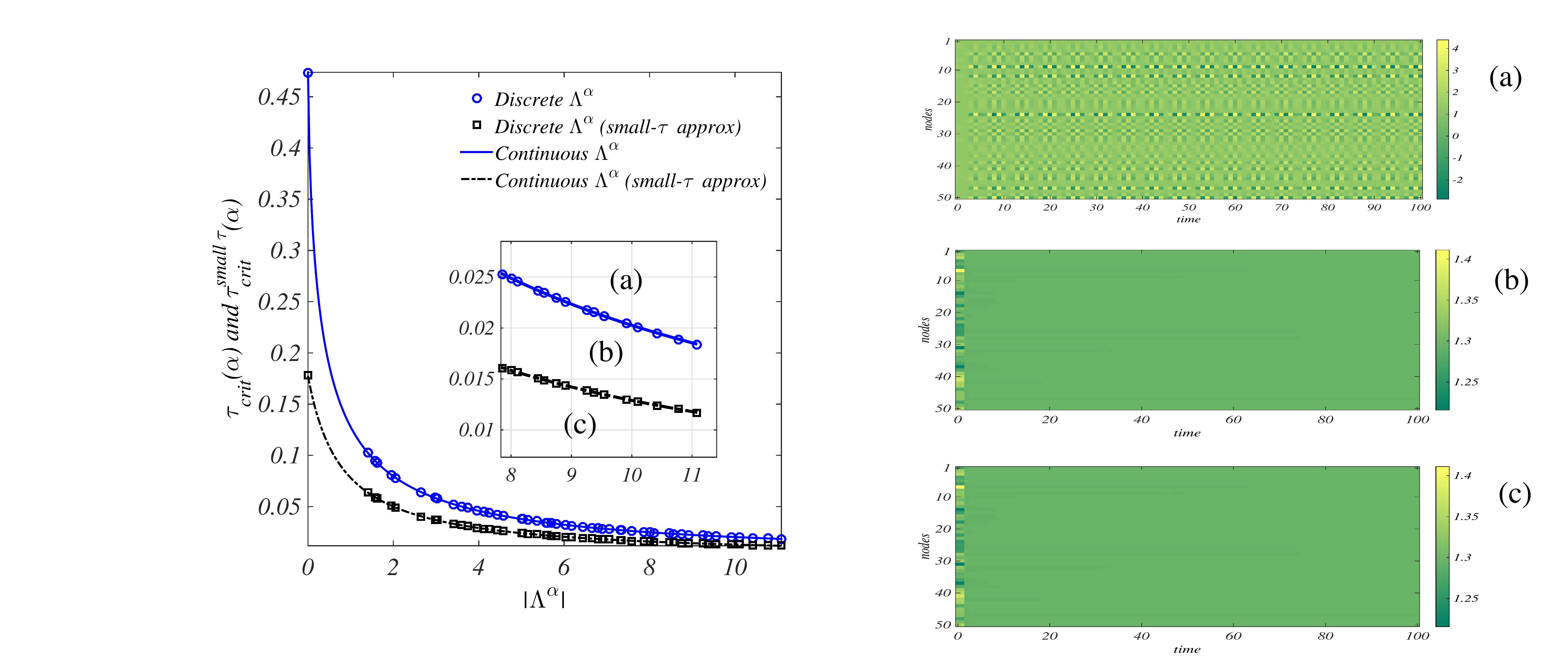} 
	\caption{Critical delay. Left panel : Computation of the critical delay Eq.~\eqref{eq:exact-critical-value-of-time-delay} (blue curve) and with the small-$\tau$ approximation (see Eq.~\eqref{eq:concmargdelsmalltau} and text below) (black curve). The inset shows a zoom for large $\lvert \Lambda^{\alpha}\rvert$, allowing to appreciate the too much conservative stability condition obtained using the small-$\tau$ approximation, even for such small values. Indeed we have $\tau_\mathrm{crit}^{\mathrm{small}\,\tau} = 1.17\cdot 10^{-2}<\tau_\mathrm{crit} = 1.84\cdot 10^{-2}$. Right panel: possible emergence of an oscillatory pattern according to the value of $\tau$. Each subplot represents the evolution of the inhibitor level over time in every node of the network, top (a) corresponds to $\tau =2.00\cdot 10^{-2}$, middle (b) corresponds to $\tau =1.50\cdot 10^{-2}$, lower (c) corresponds to $\tau =1.00\cdot 10^{-2}$. Observe that $1.00\cdot 10^{-2} < \tau_\mathrm{crit}^{\mathrm{small}\,\tau} <1.50\cdot 10^{-2} <\tau_\mathrm{crit} <  2.00\cdot 10^{-2}$, the numerical simulations confirm thus the general theory developed above that asses the emergence of patterns only in case (a), while from the small-$\tau$ approximation one would have expected patterns also in case (b). The parameters $a,b,c,d$ of the Mimura-Murray model have been fixed in Section \ref{sec:Mimura-MurrayModel}. The diffusion coefficient of the activator is $D_u = 0.01$, while $D_v$ is chosen below its the critical value thus excluding stationary patterns: $D_v = 0.9 D_\mathrm{crit} D_u$. The system is initialized with a random perturbation about the homogeneous equilibrium for every node, each one assumed to be a constant function over $-\tau \leq t \leq 0$ generated with a normal law $\mathcal{N}(0,1/25)$. The underlying network is a Watts-Strogatz one made by $50$ nodes, average degree $\langle k \rangle = 6$ and probability to rewire a link $p = 0.3$. The model has been numerically integrated using Matlab's DDE23 solver.}
	\label{fig:comparisonOfCriticalTau}
\end{figure}

In Fig.~\ref{fig:comparisonOfCriticalTau} we report some numerical results to confirm the findings of this section. On the left panel we report the critical time-delay given above while on the right panel we show the numerical integration of the Mimura-Murray for three set of parameters, in one case the time-delay is above the critical value and thus patterns are present (case (a) in the figure) while in the remaining two the time-delay is lower than the critical one and thus patterns cannot develop as clearly visible in panel (b) and (c). The reason for panel (b) will be clear later on once we  present the small-delay approximation. The remaining model parameters  have been fixed in Section~\ref{sec:Mimura-MurrayModel}, the diffusion coefficient of the activator is $D_u = 0.01$, while $D_v$ is now chosen below its the critical value in order to exclude stationary patterns: $D_v = 0.9 D_\mathrm{crit} D_u$. 

\begin{figure}[h]
	\centering
		\includegraphics[width=0.32\textwidth]{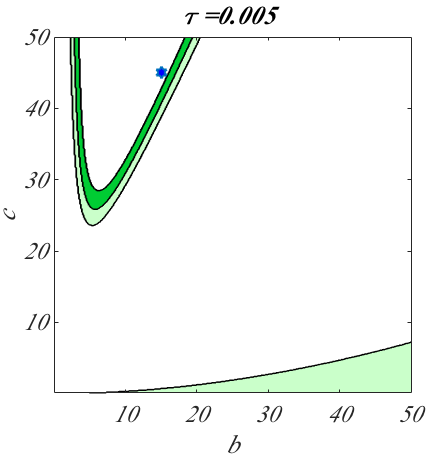} \hfill
		\includegraphics[width=0.32\textwidth]{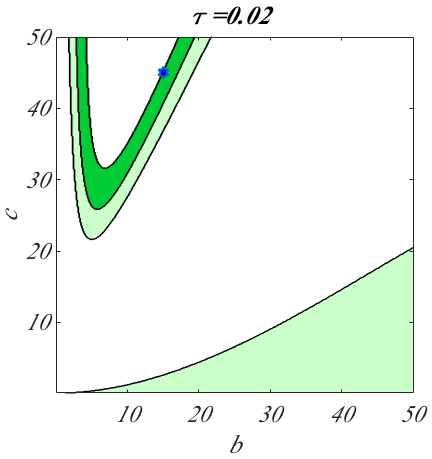} \hfill
		\includegraphics[width=0.32\textwidth]{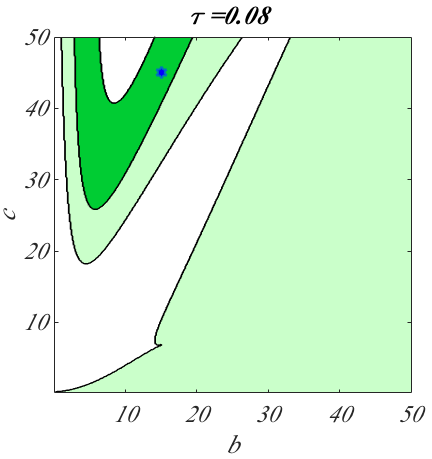} \hfill
	\caption{Turing domain (darker green region) of the system, for three different values of the delay parameter, increasing from left to right: $\tau \in \left\{ 0.005\, , \,  0.02 \, , \,  0.08 \right\}$. The  shaded region (both tones) corresponds to the domain in the $b-c$ parameter space for which the system with diffusion is unstable. The lower (lighter green) region corresponds to the instability domain of the homogeneous system, and is thus not part of the Turing domain. The remaining upper region  is thus the actual Turing domain of the system and it increases with $\tau$. Note that we are actually only interested by the darker shaded region, which respects the roles of activator and inhibitor of $u$ and $v$. The values of $b$ and $c$ used in the document are still marked by a blue star, and $a$ and $b$ have also remained unchanged. The network is still the same as in  Fig.~\ref{fig:comparisonOfCriticalTau}, with $\Lambda^{50} = -11.09$, as are the values of $D_u$ and $D_v=0.9D_\mathrm{crit}D_u$. The delay-induced Turing region of this figure is thus exclusively related to oscillations.} 
	\label{fig:TuringDom}
\end{figure}

How the time-delay impacts the shape the Turing domain by rendering possible an oscillatory behavior of the system can be seen on Fig~\ref{fig:TuringDom}. Due to the choice of $D_v$ below its critical value $D_\mathrm{crit}D_u$, we do not expect stationary patterns. The existence of a Turing domain is purely delay-driven.

\section{A small time-delay approximation}
\label{sec:small_time_delay_first_order_u_v}

In some applications, even if the delay is present, it can be considered small with respect to the natural time scale of the model and thus considered as a small parameter. For instance in~\cite{SenetAl2009}, in the framework of a reaction-diffusion system on a continuous domain where the delay impacts only the reaction term, authors developed the delayed concentrations and the reaction terms up to first order into the delay. A similar approach will be used in the following and the obtained results compared with the exact ones obtained using the Lambert $W$--function.

Assuming a small delay $\tau$, we develop $u(t-\tau)$ and $v(t-\tau)$ into power series of $\tau$ and we retain only the first order terms:
\begin{equation}
	w(t-\tau) = w(t)-\tau \frac{\partial w(t)}{\partial t} \quad\mbox{for} \quad w=u_\tau \mbox{ or } v_\tau\, . 
\end{equation}
Inserting this approximation in the reaction-diffusion model~\eqref{eq:PDS} with $\tau_r=\tau_d=\tau$, and developing also $f$ and $g$, still at first oder, we get for all $i=1,\ldots,n$:
\begin{align}
	\frac{\partial u_i}{\partial t} &= f(u_i,v_i)+f_{u_\tau}(u_i,v_i)(-\tau \frac{\partial u_i}{\partial t})
	+f_{v_\tau}(u_i,v_i)(-\tau \frac{\partial v_i}{\partial t})
	+D_u \sum_j L_{ij}u_j-D_u\tau \sum_j L_{ij}\frac{\partial u_j}{\partial t}\\
	\frac{\partial v_i}{\partial t} &= g(u_i,v_i)+g_{u_\tau}(u_i,v_i)(-\tau \frac{\partial u_i}{\partial t})
	+g_{v_\tau}(u_i,v_i)(-\tau \frac{\partial v_i}{\partial t})
	+D_v \sum_j L_{ij}v_j-D_v\tau \sum_j L_{ij}\frac{\partial v_j}{\partial t}\, ,\nonumber
\end{align}
and reordering terms we obtain:
\begin{align}
	\frac{\partial u_i}{\partial t} \left(1+\tau f_{u_\tau}(u_i,v_i)\right) +\tau f_{v_\tau}(u_i,v_i)\frac{\partial v_i}{\partial t}+D_u\tau \sum_j L_{ij}\frac{\partial u_j}{\partial t}&= f(u_i,v_i)+D_u \sum_j L_{ij}u_j\\\frac{\partial v_i}{\partial t} \left(1+\tau g_{u_\tau}(u_i,v_i)\right)+\tau g_{v_\tau}(u_i,v_i)\frac{\partial v_i}{\partial t}+D_v\tau \sum_j L_{ij}\frac{\partial v_j}{\partial t} &= g(u_i,v_i)+D_v \sum_j L_{ij}v_j\, .\nonumber
\end{align}

Let once again $(\hat{u},\hat{v})$ be an homogeneous stable fixed point, ($f(\hat{u},\hat{v})=g(\hat{u},\hat{v})=0$), consider  small perturbations $\delta u_i = u_i-\hat{u}$ and $\delta v_i = v_i - \hat{v}$ and linearize about the latter:
\begin{align}
	(1+\tau f_{u_\tau})\frac{\partial \delta u_i}{\partial t} + \tau f_{v_\tau}  \frac{\partial \delta v_i}{\partial t} + \tau D_u \sum_j L_{ij} \frac{\partial \delta u_j}{\partial t} &= f_{u_\tau} \delta u_i + f_{v_\tau} \delta v_i + D_u \sum_j L_{ij} \delta u_j 
	\\
	\tau g_{u_\tau} \frac{\partial \delta u_i}{\partial t} + (1+\tau g_{v_\tau} ) \frac{\partial \delta v_i}{\partial t} +\tau D_v \sum_j L_{ij} \frac{\partial \delta v_j}{\partial t}&= g_{u_\tau} \delta u_i + g_{v_\tau} \delta v_i + D_v \sum_j L_{ij} \delta v_j \nonumber
\end{align}
where the derivatives are now evaluated at the steady state. Using the eigenbasis for the Laplacian matrix $\phi ^\alpha _i$ to decompose the small perturbations  $\delta u_i$ and $\delta v_i$, we eventually come to the characteristic equation:
\begin{equation}
	\det 
	\begin{pmatrix}
		\lambda^\alpha (1+\tau f_{u_\tau}+\tau D_u \Lambda^\alpha) -f_{u_\tau}-D_u\Lambda^\alpha &
		(\tau \lambda^\alpha - 1) f_{v_\tau} \\
		(\tau \lambda^\alpha - 1) g_{u_\tau} &
		\lambda^\alpha (1+\tau g_{v_\tau}+\tau D_v \Lambda^\alpha) -g_{v_\tau}-D_v\Lambda^\alpha 
	\end{pmatrix} = 0, \quad \alpha = 1,\ldots,n\, .
\end{equation}
Solving for $\lambda^\alpha$ and keeping only terms up to first order in $\tau$, we get for each mode a quadratic equation: 
\begin{equation}
	\label{eq:quadlambda}
	(\lambda^\alpha)^2 (\tau \tr \mathbf{J^\alpha}+1) 
	-\lambda^\alpha (2\tau \det \mathbf{J^\alpha} +  \tr \mathbf{J^\alpha})+\det \mathbf{J^\alpha}= 0\, .
\end{equation}

To infer the stability of the homogeneous equilibrium we should consider the previous equation with $\alpha=1$, or equivalently $D_u = D_v=0$, i.e. absence of diffusion, recovering in this case an expression similar to Eq.~$(2.15)$ of~\cite{SenetAl2009}:
\begin{equation}
	\lambda^2 (\tau \tr \mathbf{J} +1)   
	-\lambda \left(  2\tau \det \mathbf{J} + \tr \mathbf{J}  \right)
	+ \det \mathbf{J}= 0\, .
\end{equation}
Let us define
\begin{equation}
	S = \frac{2\tau \det \mathbf{J} +\tr \mathbf{J}}{\tau \tr \mathbf{J} +1} \qquad \mbox{and} \qquad P = \frac{\det \mathbf{J}}{\tau \tr \mathbf{J} +1}\, ,
\end{equation}
then the homogenous equilibrium is stable if $P>0$ and $S<0$. This condition determines a time-delay that cannot be exceeded in order to maintain stability of the system
\begin{equation}
	\left\{
	\begin{array}{l}
		1+ \tau \tr \mathbf{J} >0\\
		\tr \mathbf{J} + 2 \tau  \det \mathbf{J} <0
	\end{array}
	\right.
	\quad \iff \quad
	\tau  < \min 
	\left\{
	\frac{-1}{\tr \mathbf{J}} ,
	\frac{-\tr \mathbf{J}}{2 \det \mathbf{J}}
	\right \}
	\, .
\end{equation}

Observe that the minimum of the two is $-1/\tr \mathbf{J}$ if the eigenvalues of $\mathbf{J}$ are real (or if their imaginary part is not too large).

Let us now take back into account the diffusion and determine a critical time-delay for every mode $\alpha >1$ for which there are Turing patterns provided the homogeneous equilibrium is stable. Let us define $S^\alpha $ and $P^\alpha$ for $\alpha$ such that the denominator is nonzero:
\begin{equation}
	S^\alpha = \frac{2\tau \det \mathbf{J^\alpha} +\tr \mathbf{J^\alpha}}{\tau \tr \mathbf{J^\alpha} +1} \qquad \mbox{and} \qquad P^\alpha = \frac{\det \mathbf{J^\alpha}}{\tau \tr \mathbf{J^\alpha} +1}\, .
\end{equation}

Let us now discuss the different cases.
\begin{itemize}
	\item If $\tau \tr \mathbf{J^\alpha} +1 >0$, \emph{i.e.}  $\tau<-1/\tr \mathbf{J^\alpha}$, the stability condition $S^\alpha<0\ ,\  P^\alpha>0$ reduces to $2\tau \det \mathbf{J^\alpha}+\tr \mathbf{J^\alpha}>0$, that is to say  $\tau < -\tr \mathbf{J^\alpha}/(2\det \mathbf{J^\alpha})$. 
	\item If $\tau \tr \mathbf{J^\alpha} +1 <0$, \emph{i.e.} $\tau>-1/\tr \mathbf{J^\alpha}$, then $P^\alpha>0 $ implies $\det \mathbf{J}^\alpha<0 $ which cannot hold by the stability  assumption without delay. The  condition $S^\alpha<0\ ,\  P^\alpha>0$ is never satisfied. This means that stability is never recovered by increasing the delay past $-1/\tr \mathbf{J^\alpha}$.
\end{itemize}

Observe that if $\tau \tr \mathbf{J^\alpha} +1 =0$, the quadratic equation~\eqref{eq:quadlambda} reduces to a first order equation whose solutions is 
\begin{equation}
	\lambda^\alpha = \frac{-\det \mathbf{J^\alpha}\tr \mathbf{J^\alpha}}{2 \det \mathbf{J^\alpha} - (\tr \mathbf{J^\alpha})^2 }\, ,
\end{equation}
which is negative if the eigenvalues of $\mathbf{J^\alpha}$ are real, or if their real part is larger than the imaginary part in absolute value.

In conclusion the critical delay, in the approximation of small delay, corresponding to a given mode $\alpha$ is given by:
\begin{equation}
	\label{eq:concmargdelsmalltau}
	\tau_\mathrm{crit}^{\mathrm{small}\,\tau} (\alpha) = \min 
	\left\{
	\frac{-1}{\tr \mathbf{J^\alpha}} ,
	\frac{-\tr \mathbf{J^\alpha}}{2 \det \mathbf{J^\alpha}}
	\right \}\, ,
\end{equation}
and the delay margin reads $\tau_\mathrm{crit}^{\mathrm{small}\,\tau}  = \min_{\alpha}\left\{\tau_\mathrm{crit}^{\mathrm{small}\,\tau}(\alpha) \ , \ 1<\alpha\leq n\right\}$.

The results presented in Fig.~\ref{fig:comparisonOfCriticalTau} clearly show that the small-$\tau$ approximation is not precise enough to predict the patterns onset, in fact the latter would have predicted the emergence of patterns for parameters values as in the case (b) $1.50\cdot 10^{-2} >\tau_\mathrm{crit}^{\mathrm{small}\,\tau} =1.17\cdot 10^{-2}$, while this is not the case because the complete theory shows that the critical time-delay is larger than the used one, $1.84\cdot 10^{-2}=\tau_\mathrm{crit} >1.50\cdot 10^{-2}$.

\section{Role of the network}
\label{sec:condition_on_the_Laplacian_eigenvalues}

In section \ref{sec:StationaryTuringPatterns}, we showed that stationary patterns can emerge if suitable conditions involving the reaction part, the delay term and the Laplacian eigenvalues, hold true. The aim of this section is to provide some more details about the role the network, upon which the dynamical system evolves, can play in the emergence of patterns, if any. 

The first observation is that, whatever the reaction part is, a large enough network, i.e. composed by many nodes, can impose an oscillatory behavior. Indeed, for  $|\Lambda^\alpha|$ large with respect to the magnitudes of the elements of $\mathbf{J}$, we get
\begin{align}
\label{eq:nu12forlargeLambda}
\mu_{1,2}^\alpha  & = \frac{1}{2}  \left(  \tr \mathbf{J} + (D_u + D_v) \Lambda^\alpha \pm \sqrt{(\Lambda^\alpha)^2 (D_u-D_v)^2 + 2 \Lambda^\alpha(D_u -D_v) (f_{u_\tau}-g_{v_\tau} ) + (\tr \mathbf{J})^2 -4 \det \mathbf{J}   } \right) \\
&\approx \frac{1}{2} \left(    D_u + D_v\mp |D_u-D_v|  \right) \Lambda^\alpha=  D_{u,v}\Lambda^\alpha\, .
\end{align}
The image by $W_0$ of the negative real numbers $\tau D_u \Lambda^\alpha$ and $\tau D_v \Lambda^\alpha$ will have positive real part if $|\Lambda^\alpha|$ is sufficiently large for some $\alpha$, that is $\tau D_w \Lambda^n<-\frac{\pi}{2}$ for $w = u,v$ and always a non zero imaginary part, responsible thus for the oscillating behavior, provided that the corresponding modes dominate the stationary ones.

Let us now transpose the condition for stationary patterns,  $\Re \tau \mu_1^\alpha>r_{osc}$ where $r_{osc}$ is defined by $\Re W_0(r_{osc})=\tau \ssup$, into an explicit condition on the Laplacian eigenvalues $\Lambda^\alpha$. More explicitly we now determine $\Lambda_{\mathrm{osc}}<0$  corresponding to a given  $r_\mathrm{osc}$, and thus conclude that if $\Lambda^n < \Lambda_\mathrm{osc}$, oscillatory modes will dominate, whereas if $\Lambda_\mathrm{osc} < \Lambda^n < \Lambda_\mathrm{sup}$ (where $\Lambda_\mathrm{sup}$ has been defined in~\eqref{eq:Lambda_sta}), stationary modes should emerge. 

Observe that the case $\Lambda_\mathrm{sup} \leq \Lambda_\mathrm{osc} \leq \Lambda^n$ where no conclusion can be drawn, can be avoided if we assume the network to be such that $|\Lambda^n|$ is at least  $|\Lambda_{\mathrm{sup}}|$.

For a sake of clarity, as already mentioned earlier, we will determine $\Lambda_{\mathrm{osc}}$ only in the case where $\mathbf{J}$ has real eigenvalues. The interested reader can find the remaining case of complex eigenvalues in section~\ref{sec:criticalnetworksize}. Under this assumption we can prove (see first remark of section~\ref{sec:computationOfStrongestOscillatoryMode}) that the eigenvalues $\mu_{1,2}^\alpha$ of $\mathbf{J}^\alpha$ are also real.

Let us thus assume there exists a stationary modes, hence implying $D_u < D_v$, then by Eq.~\eqref{eq:lambdacmplx1} the condition for the dominance of such stationary modes  reads
\begin{equation}
\label{eq:nun_Vs_taunumin}
\tau\frac{\tr \mathbf{J^n} - \sqrt{(\tr \mathbf{J^n})^2-4\det \mathbf{J^n}}}{2} > r_\mathrm{osc}\, ,
\end{equation}
namely
\begin{equation}
\label{eq:nun_Vs_taunumin1}
\begin{cases}
\tau \tr \mathbf{J^n} - 2  r_\mathrm{osc} \geq 0 & \qquad\mbox{(a)} \\
{(\tau \tr \mathbf{J^n} - 2 r_\mathrm{osc})^2} > {\tau^2((\tr \mathbf{J^n})^2 - 4\det \mathbf{J^n})} &\qquad \mbox{(b)}
\end{cases}
\end{equation}
From the definition of $\tr \mathbf{J}^\alpha$ for $\alpha=n$ we can rewrite condition Eq.~\eqref{eq:nun_Vs_taunumin1}$(a)$ as follows
\begin{equation}
\tau (\tr \mathbf{J} + (D_u+D_v)\Lambda^n) - 2  r_\mathrm{osc} \geq 0
\end{equation}
which leads to 
\begin{equation}
\Lambda^n \geq \frac{1}{\tau(D_u+D_v)}( 2r_\mathrm{osc}-\tau \tr \mathbf{J} ) =: \Lambda_{(a)}\, . 
\end{equation}
Note that $\Lambda_{(a)}$ is negative by the stability assumption of the system without diffusion ($\tau \tr \mathbf{J}> -\pi$) and the existence of stationary modes ($r_\mathrm{osc} < -\frac{\pi}{2}$).
	
Using the definition of $\det \mathbf{J}^\alpha$ for $\alpha=n$, and rearranging terms, condition Eq.~\eqref{eq:nun_Vs_taunumin1}$(b)$ can be rewritten as
\begin{equation}
\label{eq:condB}
\tau^2 D_uD_v (\Lambda^n)^2 + (\tau^2(\fDgD)-\tau(D_u+D_v)r_\mathrm{osc})\Lambda^n
+ (r_\mathrm{osc})^2-\tau \tr \mathbf{J}  r_\mathrm{osc} + \tau ^2 \det\mathbf{J} >0.
\end{equation}

Let us denote by $l_1$ and $ l_2$ the roots of the above polynomial in
$\Lambda^n$, with the choice $\Re l_1 \leq \Re l_2$. Their sum is negative
(remember that $r_{osc}<0$ and $\fDgD>0$), and their product has the sign of
the term:
\begin{equation}
\label{eq:productLam1Lam2}
(r_\mathrm{osc})^2-\tau \tr(\mathbf{J}) r_\mathrm{osc}+ \tau ^2 \det
\mathbf{J}\, , 
\end{equation}
that can be studied as a quadratic equation in $\tau$. If $(\tr
\mathbf{J})^2-4\det\mathbf{J}>0$, i.e. the eigenvalues of the homogeneous system
are real, then the above equation has two real roots
\begin{equation}
\tau_{1} = \frac{r_\mathrm{osc}}{\det \mathbf{J}} \mu_{1}\text{ and
}\tau_{2} = \frac{r_\mathrm{osc}}{\det \mathbf{J}} \mu_{2}\, .
\end{equation}
Using once again the stability assumption of the homogeneous system, we get
$\tau \mu_1 > -\frac{\pi}{2} > r_\mathrm{osc}$, so $\tau < \tau_1$ and thus
the quadratic equation~\eqref{eq:productLam1Lam2} is always positive and so
is the product $l_1 l_2$. 

Back to Eq.~\eqref{eq:condB} we compute its discriminant $\rho$ and we observe that it cannot be negative, in fact in this case the Eq.~\eqref{eq:condB} is satisfied for all $\Lambda^n$ and so is Eq.~\eqref{eq:nun_Vs_taunumin1}(b). Taking thus $\Lambda^n=\Lambda_{(a)}$ we get from the latter equation:
\begin{equation}
\label{eq:ifDiscriminantNeg}
0=(\tau \tr \mathbf{J}^n-2r_{osc})\rvert_{\Lambda^n=\Lambda_{(a)}} >  \tau^2((\tr \mathbf{J}^n)^2 - 4\det \mathbf{J}^n)\rvert_{\Lambda^n=\Lambda_{(a)}}\Rightarrow (\tr \mathbf{J}^{(a)})^2 - 4\det \mathbf{J}^{(a)})<0 \, ,
\end{equation}
where we defined $\mathbf{J}^{(a)}=\mathbf{J}^n\rvert_{\Lambda^n=\Lambda_{(a)}}$. But this is in contradiction with the assumption that the eigenvalues $\mu_{1,2}^\alpha$ are real, which is a consequence of the assumption $\mu_{1,2}\in\mathbb{R}$ as stated before and proved in section~\ref{sec:computationOfStrongestOscillatoryMode}.

To conclude we need to show how to obtain $\Lambda_{osc}$. Since we have just observed that the determinant $\rho$ of Eq.~\eqref{eq:condB} cannot be negative, the two solutions $l_1,l_2$ are real. We have shown that $l_1+l_2\leq 0$ and $l_1l_2>0$, so both solutions are (stricly) negative, and we can take $\Lambda_{osc} = \max(l_1,l_2)=l_2$. Note that with this choice, we have $\Lambda_{osc} >\Lambda_{(a)}$ and condition \eqref{eq:nun_Vs_taunumin1} (a) is also  satisfied (indeed, we have seen $\Lambda_{(a)}$ does not satisfy condition \eqref{eq:nun_Vs_taunumin1} (b) and so it is such that $l_1<\Lambda_{(a)}<l_2$ ).

\begin{figure}[htbp]
	\begin{center}
		\includegraphics[width=1\textwidth]{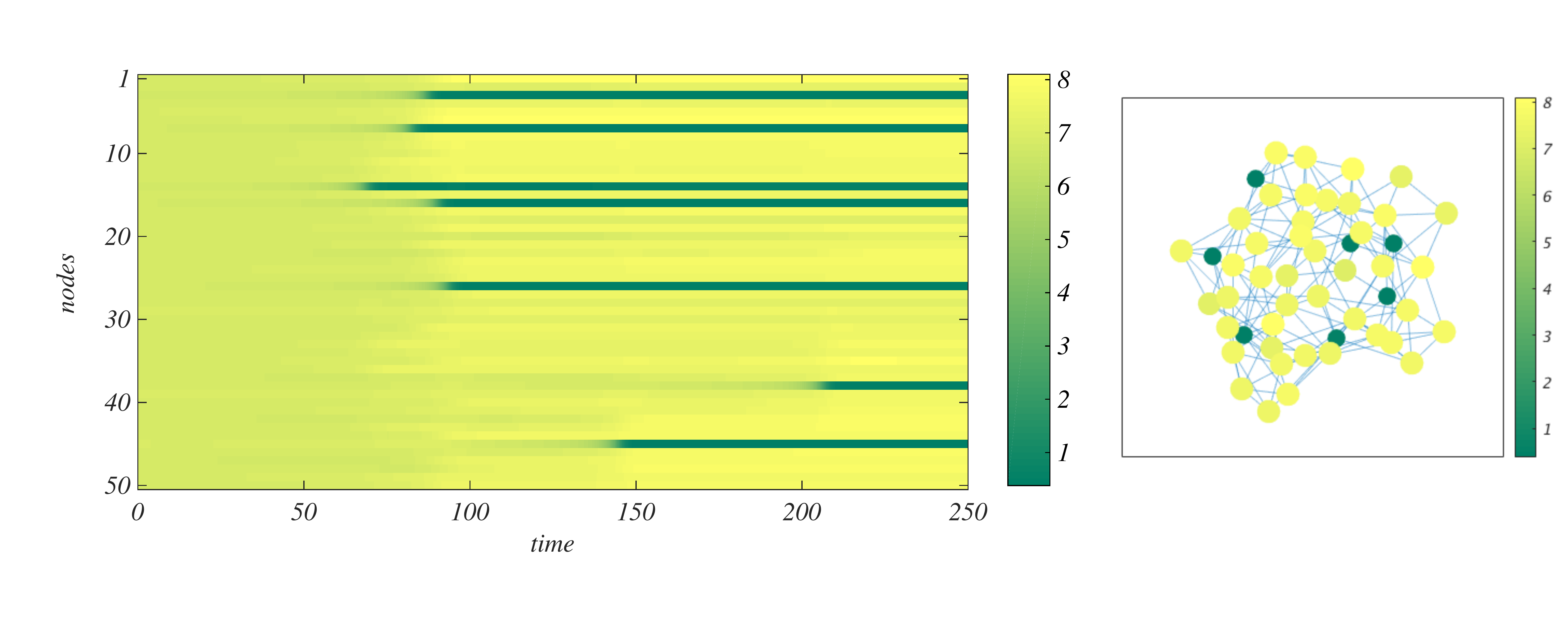} 
	\end{center}
	\caption{Stationary pattern for the Mimura-Murray interaction model. The left panel shows the evolution of the activator level in every node over time. One can observe the usual division in activator rich and poor nodes. Actually only for a few nodes the asymptotic level appears to be lower than the homogeneous equilibrium level of the activator $\hat{u}=6.83$. The right panel shows the asymptotic network configuration, the color and size of nodes is determined by the activator level (the smaller the size and the darker the color, the smaller the level of activator). The model parameters $a,b,c,d$ and the underlying network are the same as for Fig.~\ref{fig:comparisonOfCriticalTau}. The time-delay equals $\tau = 1/100$ and the diffusion coefficient of  the activator is $D_u = 0.01$, while for the inhibitor $D_v = 1.3D_{\mathrm{crit}}D_u$ to ensure stationary modes are be present. The model is initialized with a perturbation about the homogeneous equilibrium $(\hat{u},\hat{v})=(6.83,1.30)$ in every node for $-\tau \leq t \leq 0$  using constant functions, with values randomly selected from a normal distribution $\mathcal{N}(0,1/25)$. The network is such that $\Lambda_{\mathrm{osc}} =-14.60< \Lambda^n=\Lambda^{50}=-11.09<\Lambda_{\mathrm{sup}}=-8.67$ in order to let a stationary pattern develop. The spectral abscissa of the system is $0.024$. The integration was performed using a RK4 scheme adapted to deal with DDE's with constant delays.}
	\label{fig:stationaryPatterns}
\end{figure}

In Fig.~\ref{fig:stationaryPatterns} we report some numerical results, obtained using the Mimura-Murray model, to present the findings of the last sections. The numerical integration has been performed using a DDE-capable RK4 scheme, which proved to yield similar results to Matlab's DDE23 solver. The model parameters, the time-delay and the network topology have been chosen to ensure the existence of stationary patterns, namely to satisfy the conditions $\Lambda_{\mathrm{osc}} < \Lambda^n<\Lambda_{\mathrm{sup}}=-8.67$. The dynamics run on top of the same Watts-Strogatz network as in Fig.~\ref{fig:comparisonOfCriticalTau}.

\section{Role of the mobility of the species}
\label{sec:perturbation_about_Dv}
We know from the results of the previous Section~\ref{sec:existenceOfStationaryModes} that stationary modes develop when the diffusion coefficient of the inhibitor is large enough, namely $D_v>D_\mathrm{crit}D_u$, provided that there are  laplacian eigenvalues negative enough in order to let the stationary dispersion relation reach positive real part. The goal of this section is to highlight the role of $D_v$ in the onset of patterns once it gets past, but still close to the critical value, allowing us to carry out a perturbative analysis of the spectrum. 

Let us thus write 
\begin{equation}
\label{eq:Dcrit_plus_epsilon}
D_v = D_\mathrm{crit} D_u + \epsilon\, ,
\end{equation} 
with $\epsilon$ our perturbation parameter and where $D_\mathrm{crit}$ is given by expression~\eqref{eq:Dcrit}. Inserting the previous relation in the characteristic equation~\eqref{eq:characPDS}, we get that $\lambda^\alpha(\epsilon) e ^{\tau \lambda^\alpha(\epsilon)}$ is an eigenvalue of 
\begin{equation}
\label{eq:M0plusM1}
{\mathbf{J} + \begin{pmatrix}
D_u & 0 \\ 0 & D_\mathrm{crit} D_u
\end{pmatrix} \Lambda^\alpha }+ \epsilon {\begin{pmatrix}
0 & 0 \\ 0 & 1 
\end{pmatrix} \Lambda^\alpha}=:\mathbf{M_0}+\epsilon \mathbf{M_1}\, ,
\end{equation}
where the matrices $\mathbf{M_i}$, $i=1,2$ have been defined by the latter equality. Assuming $\mathbf{M_0}$ has distinct eigenvalues, we can thus use an extension of the Bauer-Fike theorem in the same fashion as done in \cite{asllanietAl2014}, and obtain the following correction to the spectrum:
\begin{equation}
\label{eq:dispersionRelationPerturbation}
\left.\lambda^\alpha( \epsilon) e ^{\tau \lambda^\alpha (\epsilon)} \right|_k= 
\left.\lambda^\alpha(0) e ^{\tau \lambda^\alpha (0)} \right|_k+ \epsilon \frac{(\mathbf{U_0} \mathbf{M_1} \mathbf{V_0})_{kk}}{(\mathbf{U_0V_0})_{kk}} 
+ \epsilon^2 \frac{(\mathbf{U_0} \mathbf{M_1} \mathbf{V_1})_{kk}}{(\mathbf{U_0V_0})_{kk}} + \mathcal{O}(\epsilon^3)
\end{equation}
where the indexes $k = 1,2$ denotes which of the two eigenvalues of $\mathbf{M_0} +\epsilon \mathbf{M_1}$ is considered, the matrix $\mathbf{V_0}$ (respectively $\mathbf{U_0}$) contains the right (respectively left) eigenvectors of $\mathbf{M_0}$ in columns (respectively rows). The matrix $\mathbf{V_1}$ is given by $\mathbf{U_0 V_1} = \mathbf{C_1}$ where 
\begin{equation}
(C_1)_{ij} = \begin{cases}
\frac{-(\mathbf{U_0} \mathbf{M_1} \mathbf{V_0})_{ij}}{\lambda_i (0)- \lambda_j(0)} & i \neq j\\
0                                                                          & i = j
\end{cases}, 
\end{equation}
with $\lambda_k (0)$, $k=1,2$, are the eigenvalues of $\mathbf{M_0}$, ordered such that $\Re \lambda_1 (0)\leq \Re \lambda_2 (0)$.

To conclude our analysis we will apply the theory developed above, in particular we will prove that the rightmost real positive characteristic root is larger than the rightmost non-real characteristic root, hence determining the emergence of stationary patterns.

For $\epsilon=0$ we have $D_v=D_{crit}D_u$ and thus~\cite{NakaoMikhailov2010} the rightmost eigenvalue $\mu_2^\alpha$ has zero value and thus $\left.\lambda^\alpha (0)e^{\tau \lambda^\alpha (0)}\right|_2= 0$. So from Eq.~\eqref{eq:dispersionRelationPerturbation} we get:
\begin{equation}
\label{eq:dispersionRelationPerturbationStat}
\ssup(\epsilon)= \frac{1}{\tau} W_0\left(\tau(\epsilon c_1 + \epsilon^2 c_2 + \mathcal{O}(\epsilon^3))\right)
\end{equation}
where the coefficients $c_1,c_2$ are computed according to~\eqref{eq:dispersionRelationPerturbation}. 

To get rid of the presence of $W_0$ in the above equation, we introduce the series expansion of the Lambert $W$--function~\cite{CorlessEtAl1996}:
\begin{equation}
\label{eq:maclaurinLambertWfunction}
W_0(z) = \sum\limits_{n=1}^{\infty} (-1)^{n+1} z^n = z-z^2+\frac{3}{2} z^3 + \mathcal{O}(z^4)\, ,
\end{equation}
the above series being valid only for $|z| < \frac{1}{e}$ where it converges. Expanding thus up to second order, the dispersion relation~\eqref{eq:dispersionRelationPerturbationStat} becomes
\begin{equation}
\label{eq:dispRelStaSeries}
\ssup(\epsilon)=\epsilon c_1 + \epsilon^2 (c_2 - \tau c_1^2) + \mathcal{O}(\epsilon^3),\quad\mbox{for every } \left|   \tau \epsilon c_1 + \tau \epsilon^2 c_2 + \mathcal{O}(\epsilon^3)   \right| < \frac{1}{e}\, .
\end{equation}

For the oscillatory characteristic root we get~\footnote{Let us observe that if $\mu_{1,2}$ are complex then to define $\snonreal$ one has to take the real part of $W_0$.}, still from~\eqref{eq:dispersionRelationPerturbation}
\begin{equation}
\label{eq:dispersionRelationPerturbationOsc}
\snonreal(\epsilon)= \frac{1}{\tau} W_0\left(\tau(d_0 +\epsilon d_1 + \epsilon^2 d_2 + \mathcal{O}(\epsilon^3))\right)
\end{equation}
where again $d_0,d_1\mbox{ and } d_2$ stem from~\eqref{eq:dispersionRelationPerturbation}. Observe that now there is a zeroth order term, $\tau d_0=\tau \lambda_1^n$, where $\lambda_1^n$ is computed using $D_v = D_\mathrm{crit} D_u$, i.e. $\epsilon=0$. Let  us also emphasize  that because we assume the eigenvalues of $\mathbf{J}$ to be real, then the nonreal rightmost characteristic root is obtained for $\alpha=n$. Expanding the Lambert $W$--function~\footnote{The Taylor series at a nonzero point $a$ can be computed~\cite{CorlessEtAl2007} using the expression for the successive derivatives of $W_0$, namely $W_0(x) = W_0(a) +W_0^{(1)}(a)(x-a) +W_0^{(2)}(a)\frac{(x-a)^2}{2} + \mathcal{O}(x-a)^3=W_0(a) + \frac{W_0(a)}{a(1+W_0(a))} (x-a) + \frac{W_0^2(a)(-2-W_0(a))}{a^2(1+W_0^3(a))} \frac{(x-a)^2}{2} + \mathcal{O}(x-a)^3$.} around this point $a = \tau \lambda_1^n$, we get an explicit formula
\begin{equation}
\label{eq:dispRelOscSeries}
\snonreal(\epsilon)= \frac{1}{\tau} W_0(a) + \epsilon W_0^{(1)}(a) d_1 + 
\epsilon^2 \left( W_0^{(1)}(a) d_2 + \frac{1}{2}\tau W_0^{(2)}(a)  d_1^2 \right) + \mathcal{O}(\epsilon^3)\, . 
\end{equation}

\begin{figure}[htbp]
	\centering
	\includegraphics[width =  \textwidth]{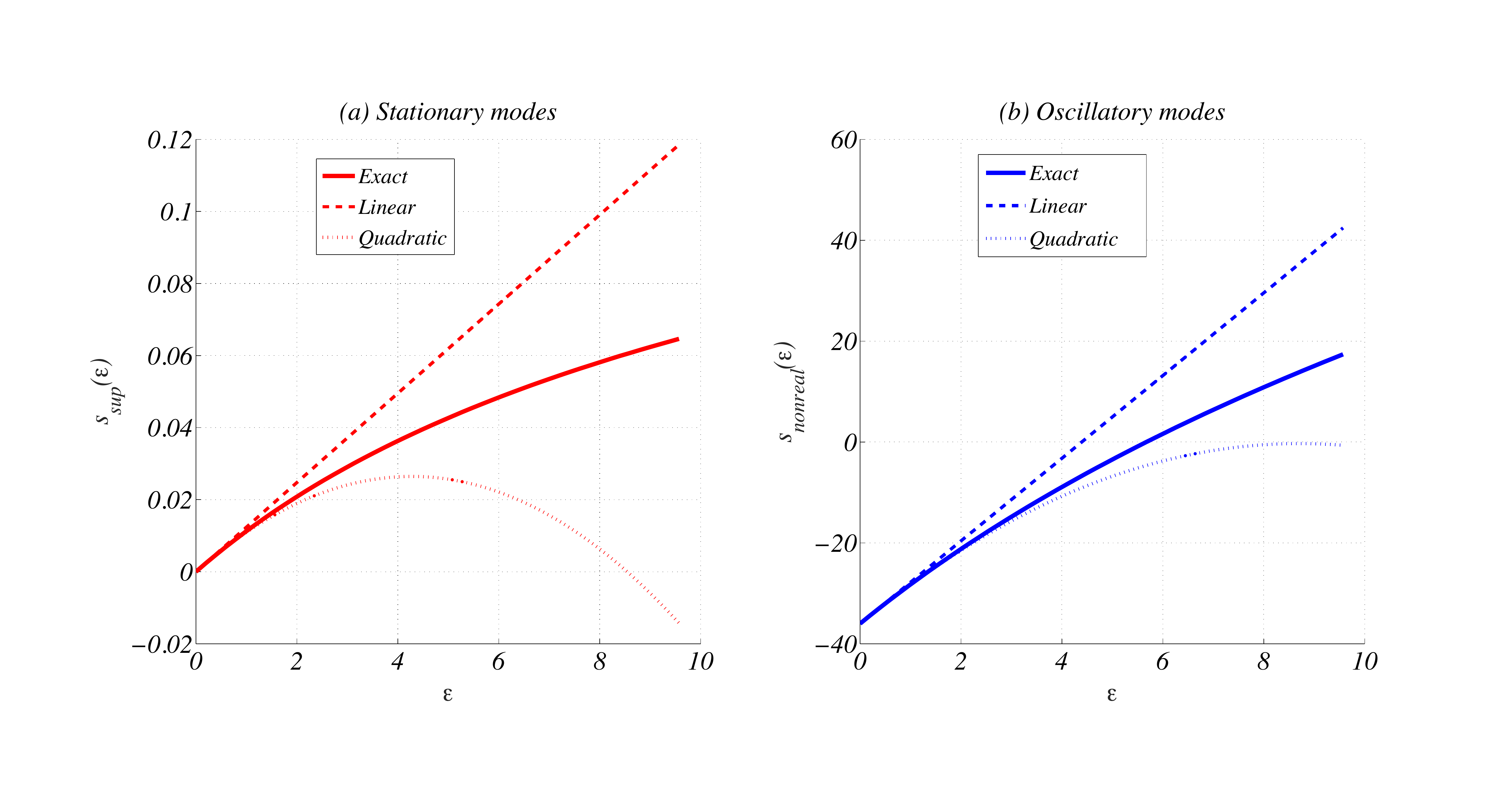} 
	\caption{Plot of the rightmost real (left panel) and nonreal (right panel)  characteristic roots of the Mimura-Murray model, for several values of $D_v=D_{crit}D_u+\epsilon$ given in terms of the perturbative parameter $\epsilon$. Except for the diffusion coefficient of the inhibitor which is varied here, all other parameters are the same as given in the caption of Fig.~\ref{fig:stationaryPatterns}. The values assumed by the real characteristic roots $\ssup(\epsilon)$ tend to be rapidly overcame by the ones taken by the oscillatory characteristic roots $\snonreal(\epsilon)$, which determine the spectral abscissa of the system when the mobility of the inhibitor is increased. In both panels the solid line denotes the exact numerical values for $\ssup(\epsilon)$ and $\snonreal(\epsilon)$, while the dashed line is the first order approximation and the dotted line the second order one.}
	\label{fig:perturbDv}
\end{figure}

Using once again the Mimura-Murray model for the reaction part, we show in Fig.~\ref{fig:perturbDv}, the results given by the previous Eqs.~\eqref{eq:dispRelStaSeries} and~\eqref{eq:dispRelOscSeries}. More precisely we report in such figure the exact rightmost real (left panel) and nonreal (right panel) characteristic root, together with the first and second order approximation as a function of $\epsilon$. We set parameters in such a way for small $\epsilon>0$ the model exhibits stationary patterns, namely $\snonreal<0<\ssup$ (see Eq.~\eqref{eq:statvsoscpatt}); we can observe that for a quite large range of values of positive $\epsilon$, stationary patterns still dominate, but as soon as the non-real rightmost characteristic root gets a positive real part, $\epsilon\sim 5.7$, then oscillatory patterns take over. Stated differently, higher mobility of the inhibitor favours the emergence of oscillatory modes. Let us observe that the perturbative scheme doesn't have the same accuracy for the two cases, hence if one wants to use this approach to determine the value of $\epsilon$ for which $\ssup(\epsilon)=\snonreal(\epsilon)$, then the order of the perturbative developments has to be high enough.

\section{Conclusions}
\label{sec:conc}

Reaction-diffusion systems on complex networks are attracting increasingly attention in the scientific community because of their diversified applications to interdisciplinary problems arising from very different disciplines. In this framework the theory of Turing instabilities plays a relevant role to explain the emergence of self-organized spatio-temporal patterns. Following a symmetry breaking instability, seeded by diffusion, a stable homogeneous fixed point of the reaction term can be destabilized by an inhomogeneous perturbation. The linear regime associated to short times is then followed by a second one where patterns possibly form due to the nonlinearities in the reaction term. Because of the former, the conditions that underly the instability process can be obtained via a standard linear stability analysis, which requires
expanding the imposed perturbation on a basis formed by the eigenvectors of the network Laplacian
matrix. 

Starting from this setting, we have  analyzed here the case where the reaction-diffusion model contains a time-delay term; both the reactions and the diffusion take some time to be realized which implies that the growth rate of the involved two species depends on their concentration in the past. The linear stability analysis is now more complicated than in the delay--free case because the characteristic roots, from which  the relation dispersion is computed, are the denumerably many solutions of a quasi-polynomial equation. To deal with this issue we use the Lambert $W$--function, allowing to cast the conditions in analytically closed form.

The dispersion relation which ultimately determines the onset of  instability, is function of three main parameters, the time-delay, the species diffusivity and the network topology through its spectrum. Dealing with analytical formulas allowed us to completely describe the inception of pattern formation as a function of the above quantities. In particular we provided conditions for the emergence of stationary Turing patterns. This is an important new result because other works in the literature involving time-delay, always associated the emergence of patterns with a Hopf-bifurcation, thus  letting only waves possibly develop. Our conditions have been given in terms of the time-delay but also of the network Laplacian spectrum, and we were able to obtain a complete overview of the relative importance of these parameters.

Our findings have been corroborated by direct numerical integration of the reaction-diffusion equations with time-delay, taking the Mimura-Murray kinetics as a representative model.

Building upon this work, further research  questions could now  be addressed, such as additionally characterizing the asymptotical  properties of the solutions to our model (shape and other properties of the patterns, boundedness), or generalizing the setting, for instance to different delays in the reactions and the diffusion, or even to node- or link-dependent (distributed) delays.

\section*{Acknowledgements}
The work of J.P., T.C., M.A. presents research results of the Belgian Network DYSCO (Dynamical Systems, Control, and Optimisation), funded by the Interuniversity Attraction Poles Programme, initiated by the Belgian State, Science Policy Office.

\appendix
\section{Analytical developments}
The aim of this appendix is to present some more details about the results presented in the main text.
 
\subsection{Computation of the rightmost nonreal characteristic root}
\label{sec:computationOfStrongestOscillatoryMode}

We hereby make two remarks, allowing us to compute the rightmost nonreal characteristic root, even in the case the Jacobian matrix of the system with or without diffusion, has nonreal eigenvalues.

\paragraph{First remark} It concerns the conditions under which the eigenvalues $\mu_{1,2}^\alpha$ are real numbers. Recalling their definition
\begin{equation*}
\mu_{1,2}^\alpha=\frac{\tr \mathbf{J^\alpha}\pm\sqrt{(\tr \mathbf{J^\alpha})^2-4\det \mathbf{J^\alpha}}}{2}\, ,
\end{equation*}
we conclude that one needs $(\tr \mathbf{J^\alpha})^2-4\det \mathbf{J^\alpha} \geq 0$, or equivalently using the Eqs.~\eqref{eq:trJadetJa}
\begin{equation}
\label{eq:condForRealMuAlpha}
(D_u-D_v)^2 (\Lambda^\alpha)^2
+2(D_u-D_v)(f_{u_\tau}-g_{v_\tau})\Lambda^\alpha 
+(\tr\mathbf{J})^2-4\det\mathbf{J}\geq 0\, .
\end{equation}

We assume that there are stationary modes, so we have $D_v > D_u$.

If the eigenvalues $\mu_{1,2}$ of $\mathbf{J}$ are real, then $(\tr\mathbf{J})^2-4\det\mathbf{J}>0$ and thus the above quadratic polynomial is always positive for negative $\Lambda^\alpha$. Hence $\mu_{1,2}^\alpha$ are real numbers.

On the other hand, if the eigenvalues $\mu_{1,2}$ of $\mathbf{J}$ are complex numbers, and still $D_v > D_u$, the quadratic polynomial is positive, and thus the eigenvalues $\mu_{1,2}^\alpha$ are real, only for $\alpha$ such that $\Lambda^\alpha\leq \Lambda_0$ where:
\begin{equation}
\label{eq:Lambda_zero}
\Lambda_0= \frac{f_{u_\tau}-g_{v_\tau}-2\sqrt{|f_{v_\tau}|g_{u_\tau}}}{D_v-D_u}\, .
\end{equation}

\paragraph{Second remark} It concerns the determination of the real and smallest negative eigenvalue $\mu_{1}^\alpha$ as a function of $\Lambda^\alpha$.

Let us consider firstly the case where the eigenvalues are real, that is $(\tr \mathbf{J^\alpha})^2-4\det \mathbf{J^\alpha} \geq 0$. Computing the derivative of $\mu_{1}^\alpha$ with respect to $\Lambda^\alpha$ and considering only its sign, we get
\begin{equation}
\label{eq:signDerivativeLHS}
\sign \frac{\mathrm{d}}{\mathrm{d}\Lambda^\alpha} \left(
\frac{\tr \mathbf{J^\alpha} - \sqrt{(\tr \mathbf{J^\alpha})^2-4\det \mathbf{J^\alpha}}}{2}
\right) =\sign\left[\frac{\mathrm{d}\tr \mathbf{J^\alpha}}{\mathrm{d}\Lambda^\alpha}   \sqrt{(\tr \mathbf{J^\alpha})^2-4\det \mathbf{J^\alpha}} - \left(
\tr \mathbf{J^\alpha} \frac{\mathrm{d}\tr \mathbf{J^\alpha}}{\mathrm{d}\Lambda^\alpha}  - 2 \frac{\mathrm{d}\det \mathbf{J^\alpha}}{\mathrm{d}\Lambda^\alpha} 
\right)\right]\, .
\end{equation}

The first term in the right hand side is positive because $\frac{\mathrm{d}\tr \mathbf{J^\alpha}}{\mathrm{d}\Lambda^\alpha}  =D_u+D_v$. Developing the second one gives
\begin{align}
- \left(
\tr \mathbf{J^\alpha} \frac{\mathrm{d}\tr \mathbf{J^\alpha}}{\mathrm{d}\Lambda^\alpha}  - 2 \frac{\mathrm{d}\det \mathbf{J^\alpha}}{\mathrm{d}\Lambda^\alpha} 
\right) &=-(\tr \mathbf{J}+(D_u+D_v)\Lambda^\alpha)(D_u+D_v)+2(\fDgD)+4(D_uD_v)\Lambda^\alpha\\
&= -\tr \mathbf{J} (D_u+D_v)+2(\fDgD)-(D_u-D_v)^2\Lambda^\alpha\, ,
\end{align}
and because each term is positive it follows that
\begin{equation}
\label{eq:signDerivativeLambda1alpha}
\frac{\mathrm{d}}{\mathrm{d}\Lambda^\alpha} \left(
\frac{\tr \mathbf{J^\alpha} - \sqrt{(\tr \mathbf{J^\alpha})^2-4\det \mathbf{J^\alpha}}}{2}
\right) >0\, .
\end{equation}

This means that $\mu_1^\alpha$ is an increasing function of $\alpha$ and thus the smallest value is reached for $\alpha=n$, that is the one associated to the smallest eigenvalue $\Lambda^n$.

Let us then consider the remaining case, namely $(\tr \mathbf{J^\alpha})^2-4\det \mathbf{J^\alpha} <0$. Because of the above, this holds true only if $\mu_{1,2}$ are complex and for indexes $\alpha$ corresponding to $\Lambda^\alpha >\Lambda_0$, given by Eq.~\eqref{eq:Lambda_zero}. 

For this range of topological eigenvalues, $\Lambda_0 < \Lambda^\alpha < 0$,  we need to look numerically for the maximum of  $\xi (x,y)$ with $x=\Re \tau \mu_1^\alpha = \tau \frac{\tr \mathbf{J^\alpha}}{2} $ and $y=\Im\tau \mu_1^\alpha  = -\tau \frac{\sqrt{4\det \mathbf{J^\alpha}-(\tr \mathbf{J^\alpha})^2}}{2}$.

To conclude the second remark, we find the spectral abscissa related to nonreal characteristic roots, where this time - by opposition to the main body - we include the case there exist nonreal eigenvalues of $\mathbf{J}^\alpha$ for a range of $\Lambda^\alpha$ values we have determined using the first remark: 
\begin{equation}
\label{eq:lambdacmplx2}
\snonreal=\frac{1}{\tau} \max \left\{
\Re W_0\left(\tau\frac{\tr \mathbf{J}^n - \sqrt{(\tr \mathbf{J}^n)^2-4\det \mathbf{J}^n}}{2}\right),
\max_{\Lambda_0< \Lambda^\alpha <0}
\Re W_0 \left( \tau\frac{\tr \mathbf{J^\alpha} - \sqrt{(\tr \mathbf{J^\alpha})^2-4\det \mathbf{J^\alpha}}}{2}\right)
\right\},
\end{equation}
where $\Lambda_0$ is given by Eq.~\eqref{eq:Lambda_zero}.

Note that so far in this section, we have assumed that stationary modes exist, implying that $D_v >D_u$. If to the contrary, the only characteristic roots with positive real part are nonreal, the above reasoning can be followed in a straightforward way, if we notice that nothing changes in the case that $D_{crit}D_u>D_v\geq D_u$. And if $D_u>D_v$, then it is easy to see that the discriminant of \eqref{eq:condForRealMuAlpha} is positive, and that both solutions, say $\Lambda_1\leq \Lambda_2$, are negative. So we are in the case of nonreal eigenvalues above only if $\Lambda_1<\Lambda_\alpha<\Lambda_2$.

\subsection{Role of the network in the case of nonreal eigenvalues}
\label{sec:criticalnetworksize}

Here we compute a critical topological eigenvalue, that is the largest eigenvalue of the Laplacian that still permits stationary modes to develop, in the case of nonreal eigenvalues. Recall that a necessary condition for stationary modes is $D_v>D_v$.

The first case, where $\mathbf{J}$ has real eigenvalues was treated in the main body. According to 
the first remark of section \ref{sec:computationOfStrongestOscillatoryMode}, and since $D_v>D_u$ there are still two possible options when the eigenvalues $\mu_{1,2}$ are not real: the largest $|\Lambda^\alpha|$ is greater than, or smaller than $|\Lambda_0|$. In other words, there exist real $\mu_1^\alpha$ or not. 
\begin{itemize}
\item In the second case, we assume $\mu_1,\mu_2$ are not real, but $\mu_1^n$ is real and satisfies the necessary condition $r_{osc}<\Re \mu_1^n$ and we proceed as before. It is  easy to see that in this case the  sign of  the product $l_1 l_2$, the two solutions of \eqref{eq:condB}, given by  the sign of \eqref{eq:productLam1Lam2},  is positive.
	
If the discriminant $\rho$ of \eqref{eq:condB} were negative,  the inequality \eqref{eq:ifDiscriminantNeg} would apply or equivalently, \eqref{eq:nun_Vs_taunumin1}(b) and we would have :
\begin{equation}
0 >  \tau^2((\tr \mathbf{J^n})^2 - 4\det \mathbf{J^n})\qquad \mbox{for} \qquad \Lambda^n = \Lambda_{(a)}.
\end{equation}
Following from the first remark of section~\ref{sec:computationOfStrongestOscillatoryMode}, we would have $\Lambda_0 \leq  \Lambda_{(a)} $. From condition~\eqref{eq:nun_Vs_taunumin1}(a), $\Lambda_{(a)} \leq \Lambda^n$, and so  $\Lambda_0 \leq  \Lambda^n $ which would mean $(\tr \mathbf{J^n})^2 - 4\det \mathbf{J^n}$ is negative, a contradiction with the hypothesis $\mu_1^n \in \mathbb{R}$.
		
So the discriminant $\rho$ of of \eqref{eq:condB} has to be positive, hence the two solutions $l_1,l_2$ of Eqn~\eqref{eq:condB} are reals. Since their product was shown to be positive and their sum negative, we have $l_1<l_2<0$. Assuming that $\mu_1^n$ indeed corresponds to the rightmost non-real characteristic root (and not one of the nonreal $\mu_1^\alpha$ associated to an eigenvalue of the Laplacian such that $\Lambda_0 < \Lambda^\alpha < 0$), we have $\Lambda_{\mathrm{osc}}=\max (l_1,l_2) = l_2$.

In order to verify the last assumption, we have to proceed numerically following the explanation given in the third case below.

	\item In the third case, we assume the  eigenvalue with smallest real part $\mu_1^n$ is not real, but that the necessary condition $\Re \tau \mu_1^n > r_\mathrm{osc}$ is still fulfilled. We can compute the boundary of the domain in the complex plane given by $\Re W_0(z) = \ssup$. One half of the boundary (above the real axis)  is determined by the points $x+iy$, that we find the following way. For every $x \in [r_\mathrm{osc},0]$,  we solve   system \eqref{eq:xyFunctionXiEta} for $y$ and $\eta$,   with $\xi = W_0((\tau \nu_2)_{\mathrm{max}}$ and knowing $0\leq \eta \leq \pi$. (The second half of the boundary is symmetric, and corresponds to negative $\eta$ values).
	The critical value of $\Lambda_{\mathrm{osc}}$ is given by the intersection of this boundary with the curve determined by  $\tau \nu_1^\alpha$ in the complex plane, when $\Lambda^\alpha$ is continuously varied in $\mathbb{R}^-$.
\end{itemize}

\subsection{Computation of the delay margin}
\label{sec:DelayMarginComplexCase}

Here we present the computation of the delay margin in the case of nonreal eigenvalues of the Jacobian matrix. 

\begin{description}
	\item[Without diffusion]
	If the eigenvalues of $\mathbf{J}$ are nonreal numbers, the conditions for the stability are:
	\begin{align}
	\tau \Re \mu_1 &< \frac{-\pi}{2 } & (a) \nonumber\\
	\tau \left| \Im \mu_1 \right| &< \eta \cos \eta \quad \mbox{with} \quad \frac{\tau \tr \mathbf{J}}{2}  = - \eta  \sin \eta, \ \eta \in  (0,\pi /2) & (b) \nonumber
	\end{align}
	or equivalently
	\begin{eqnarray}
	\label{eq:stanodiff}
	\tau  &<& -\pi (\tr \mathbf{J})^{-1}     \hspace{8cm}      (a) \notag\\
	\tau  &<& -2 \eta \sin\eta   (\tr \mathbf{J})^{-1} \quad \mbox{with} \quad \eta =\arctan \frac{- \tr \mathbf{J}}{2(4 \det \mathbf{J} - (\tr \mathbf{J})^2)^{1/2}}\qquad (b)
	\end{eqnarray}
	Note that $0< \eta < \pi/2$, and because over this interval $\eta \sin \eta$ is an increasing function we have $2\eta \sin \eta < \pi \sin \frac{\pi}{2} = \pi$, and the stability condition reduces to 
	\begin{equation}
	\tau  < -2 \eta \sin\eta   (\tr \mathbf{J})^{-1}\, ,
	\end{equation}
	with $\eta$ given in Eq.~\eqref{eq:stanodiff}(b).
	
	\item[With diffusion] We consider the modes $\alpha$ for which the eigenvalues of the Jacobian matrix of the system are nonreal numbers. By using our computation without diffusion above, we readily obtain
		\begin{equation}
		\label{eq:exact-critical-value-of-time-delay-nonreal-case}
		\tau <  -2 \eta \sin\eta   (\tr \mathbf{J^\alpha})^{-1}         \quad \mbox{with}
		\quad \eta =\arctan \frac{- \tr \mathbf{J^\alpha}}{2(4 \det \mathbf{J^\alpha} - (\tr \mathbf{J^\alpha})^2)^{1/2}} \ .
		\end{equation}

\end{description}



\end{document}